%% file: Esclapez_CF_MIST.tex
\documentclass[5p]{elsarticle}
\usepackage[T1]{fontenc}
\usepackage{amssymb}
\usepackage{amsmath}
\usepackage[english]{babel}
\usepackage{multirow}
\usepackage[table]{xcolor}
\usepackage{ctable}
\usepackage{array}
\usepackage{colortbl}
	
\usepackage{epsfig}			% pour inclure des images
\usepackage{graphicx}			% pour inclure des images
\usepackage{graphicx,psfrag}
\usepackage{psfrag}             		% pour remplacer des caractres sur les .eps 
\usepackage{subfig}
\usepackage{float}
\usepackage{soul}
\usepackage[final]{pdfpages}

\newcolumntype{C}[1]{>{\centering\let\newline\\\arraybackslash\hspace{0pt}}m{#1}}

\journal{Combustion and flames}

\begin{document}

%\includepdf{./Front_page/Front_page.pdf}

\begin{frontmatter}

\selectlanguage{english}

\title{A statistical model to predict ignition probability}

\author[cerfacs,SAE,LBNL]{L. Esclapez}
\author[cerfacs,SAE]{F. Collin-Bastiani}
\author[cerfacs]{E. Riber}
\author[cerfacs]{B. Cuenot}

\address[cerfacs]{CERFACS, 42 avenue G. Coriolis, 31057 Toulouse Cedex 01, France } 
\address[SAE]{Safran Aircraft Engines, Rond-Point Ren\'e Ravaud-R\'eau, 77550 Moissy Cramayel, France}
\address[LBNL]{now at Center for Computational Sciences and Engineering (CCSE), LBNL, 1 Cyclotron Road, Berkeley, CA 94720, USA}

\begin{abstract}

Ignition capability is a critical design constraint for aeronautical gas turbines. However the current trend toward overall lean burn is detrimental to the engine ignition and relight and the ignition system must be adapted to ensure a fast and reliable light-round in all circumstances. As ignition is a stochastic phenomenon, the optimization of an ignition system requires to build ignition probability maps, which is difficult and costly with either experiment or numerical simulation as both require many tests. This work proposes a model to predict the ignition probability map, knowing only flow statistics in non-reacting conditions, i.e., with only one test. The originality of the model is to construct statistics of the flame kernel trajectory, which are then combined with local flow indicators to evaluate the ignition probability at the considered sparking location. Application to a swirled burner operated in premixed, non-premixed and spray combustion modes illustrates the model concepts and demonstrates its ability to recover the experimental ignition map with good accuracy.

\end{abstract}


\end{frontmatter}

%% INTRO %%
\section{Introduction}
\label{sec_introduction}

In response to rising concerns regarding the effect of aviation emissions on the climate, the design of modern gas turbine combustors is drastically changing. Current strategies to reduce NO$_x$ and CO$_2$ emissions rely on lean combustion and optimization of the combustor design to reduce the engine weight and complexity. These new concepts raise the critical issue of high altitude relight, which is regarded as one of the more stringent constraint on the aeronautical gas turbine design. Most engines in circulation were designed using empirical correlations resulting from extensive experimental test campaigns~\cite{Lefebvre:1998}. Today high performance numerical tools are available and play an increasing role in the context of development cost saving. %BCThis paper proposes an efficient and practical numerical approach to predict ignition in aero-engines.

%As a consequence, the empirical correlations can no longer be used to evaluate ignition performances of a combustion chamber and new methodologies must be developed to guide engineering decisions, early in the development process.

%BCFollowing the early work of Lefebvre~\cite{Lefebvre:1998}, gas turbine ignition may be split in four successive steps:
%BC\begin{itemize}
%BC\item flame kernel generation by a high energy source (either spark plug or laser)~\cite{Bradley:2004,Collin-Bastiani:2018b},
%BC\item kernel growth and transition from a laminar to a turbulent flame~\cite{Abdel-Gayed:1988},
%BC\item flame propagation and stabilization on the injection system~\cite{Mastorakos:2009, Mastorakos:2017},
%BC\item flame propagation to all the injection systems until a stable combustion regime~\cite{Bourgouin:2013,Barre:2014}.
%BC\end{itemize}
%BCAll four steps must be successful in order to achieve complete ignition of a gas turbine combustor. Because each step involves processes of increasing time and space scales, they are often studied individually through experiments and numerical simulations. 

The stochastic nature %of the first three steps 
of the ignition process has been well highlighted with multiple experiments reported in the literature. Stochasticity originates from variations of the size and strength of the energy deposited by the ignition system~\cite{Kono:1984}, the turbulent flow and the reactants mixing at the sparking location~\cite{Mastorakos:2009,Sforzo:2015}, and the large-scale flow variations in the combustor~\cite{Ahmed:2007a,Cordier:2013}. Owing to these observations, ignition performances are quantified  with ignition probability $P_{ign}$ maps~\cite{Ahmed:2007a,Cordier:2013}. Conditions that maximize $P_{ign}$ are: 1) high flammability and/or rapid fuel availability at the spark location, 2) low turbulence intensity around the spark location, and 3) large-scale flow patterns allowing the flame to propagate toward the burner nozzle. The detailed analysis of the ignition process also reveals that ignition success is not solely conditioned by the local flow properties at the igniter position, but also by the flow conditions along the flame kernel trajectory after sparking.
From the numerical point of view, the transient and stochastic nature of ignition calls for the Large Eddy Simulation (LES) approach, which has been proven to accurately predict ignition in configurations representative of gas turbines~\cite{Boileau:2008, Triantafyllidis:2009, Subramanian:2010, Jones:2010}. However, although the direct prediction of the ignition probability using LES has been proven feasible~\cite{Esclapez:2015}, building a full ignition probability map is not possible due to the computational cost of tens of simulations of ignition sequences at each point of the map.

%but can seldom be used for design purposes due to the numerical cost and return time associated with LES simulations.

Rapid and computationally affordable evaluation of ignition probability maps was first proposed in the pioneering work of Birch and co-worker~\cite{Birch:1977, Birch:1981,Smith:1986} who developed a model using experimental measurements of fuel distribution and velocity. They distinguished between the $P_{ign}$ and the kernel initiation probability $P_{ker}$, which was shown to be correlated to the flammability factor $F_f$ defined as the local probability of the mixture to be in flammable conditions.
%A method was then proposed to evaluate $F_f$ based on a presumed probability density function (PDF) of the local mixture fraction and experimental measurements of the mean and fluctuating values of $z$. 
These early studies recently inspired the development of more advanced methods which can be sorted in two classes: 1) the probability is evaluated from the flow properties at the sparking location only, 2) the model tracks the spatio-temporal evolution of the ignition kernel to evaluate its chance of igniting the burner. In the first class several criteria, based on the local flammability, turbulence intensity and velocity direction, are evaluated from the non-reacting flow to evaluate the success of ignition. Stochasticity is retrieved from the analysis of multiple instantaneous flow fields which are combined as independent initial states leading to independent ignition events to construct the ignition probability~\cite{Linassier:2013, Eyssartier:2013}. These methods are computationally fast and provide a good estimation of $P_{ker}$ but they usually fail to predict $P_{ign}$, as they ignore the subsequent flame kernel evolution. Model of the second class were initiated by Wilson \emph{et al.}~\cite{Wilson:1999}, where the dispersion of a conserved scalar in simulation of the non-reacting flow is used to track possible kernel trajectories. A more recent attempt ~\cite{Richardson:2007b, Weckering:2011, Neophytou:2012} introduced the Lagrangian tracking of representative flame particles, adding artificial stochasticity to the mean flow. In both methods, the Karlovitz number was used to evaluate the occurrence of flame quenching~\cite{Soworka:2014}. 
These methods are intrinsically well suited to capture transient flame kernel motion and expansion, although they were found much sensitive to the success criteria thresholds and required multiple simulations to obtain converged statistics. Additionally, they do not use the true flow statistics along the kernel trajectories, which is known to strongly vary spatially in complex geometries. Note that all these methods stay valid only as long as the flame kernel stays small, and should not be used once it has given birth to a large turbulent flame which modifies the flow. In particular it should not be used to predict annular light-round in azimuthal combustors where burnt gas expansion greatly affect the flame propagation.% \cite{Machover:2017}. 

In this work, a reduced order model to predict the ignition probability of modern gas turbine combustors (i.e., featuring one or more recirculation zones stabilizing the flame) is proposed. In contrast with the Monte-Carlo approach used by Neophytou \emph{et al.} \cite{Neophytou:2012}, the model includes the real, local flow statistics along the kernel trajectories, which can be extracted from time-averaged non-reactive flow quantities. This allows to take into account the complexity of the flow in the combustion chamber for an improved prediction of ignition probability and of its sensitivity to the geometrical design.  The model development and test are based on experiment and simulation of a lean swirled burner operated in premixed, non-premixed and two-phase flow combustion modes~\cite{Cordier:2013,Collin-Bastiani:2018}, well representative of real gas turbine conditions and flows.  %This allows to build reliable full ignition maps and to determine the optimum location of the igniter. 
%Motivated by the results of fully resolved LES~\cite{Esclapez:2015,Collin-Bastiani:2018}, the model focuses on the prediction of ignition probability at the end of the kernel expansion (i.e. success of step 1 and beginning of step 2) to provide a spark plug location optimization.

The paper is organized as follows. Section~\ref{sec_configuration} introduces the experimental set-up and the numerical results upon which the model is developed and tested. In Section~\ref{sec_model} the Model for Ignition STatistics (MIST) is derived and in Section~\ref{sec_results} the results of MIST applied to the test configuration are presented. Finally, the model outputs and performances are discussed, and future developments are provided in the conclusion.

%% CONFIG / LES RESULTS %%

\section{Test configuration}
\label{sec_configuration}

\subsection{Experimental configuration}
\label{ssec:XPsetup}

The experimental configuration employed to evaluate the model performances was specifically designed by Cordier \emph{et al.}~\cite{Cordier:2013, Cordier:2013b} to study ignition in complex flows, representative of realistic gas turbines, first with gas only (methane) and later with liquid fuel injection ($n$-heptane)~\cite{Marrero-Santiago:2017,Collin-Bastiani:2018}. A picture of the test rig is presented in Fig.~\ref{fig:geom}(a). The burner is capable of operating in premixed ($P$), non-premixed ($NP$) and spray ($SP$) modes at two levels of swirl intensity. It is made of four major components, namely a plenum, a swirled injection system, a combustion chamber and a convergent exhaust. The flow entering the plenum is first tranquilized through three grids before entering the swirler vanes. The combustion chamber has a 100 mm side length square section and is 260 mm long. A convergent exhaust ends the combustion chamber to avoid air admission induced by the swirling flow. Finally, the injection system is composed of a central jet ($d = 4$ mm) nested within the annular swirl stream ($D_{in} = 9$ mm, $D_{ext} = 20$ mm) of the swirler, the latter consisting of 18 radially fed channels inclined by 45 degrees. In $P$ mode, both the central tube and the plenum are fed with a methane/air mixture whereas in $NP$ mode the central jet is fed with pure methane and the plenum is fed with air. In spray mode, the central jet injection tube is replaced by a simplex pressure atomizer (Danfoss, $1.46 kg/h$, $80^o$ hollow cone) fueling liquid $n$-heptane. 

All experimental operating conditions of modes $P$, $NP$ and $SP$ are summarized in Tab.~\ref{Tbl:expe_cond}. Contrary to gaseous cases, air and fuel are preheated in the $SP$ case and a leaner regime is studied. In non-reacting conditions, stereoscopic particle image velocimetry (SPIV) is used to measure the three components of velocity in a 50 mm $\times$ 67 mm field of view. Statistics of velocity are computed from 1000 images. PDA measurements were used to characterize the liquid phase in terms of droplet size and size-classified velocity. 
To measure the fuel mole fraction field, planar laser induced fluorescence (PLIF) based on acetone is used in $NP$ mode while Toluene-PLIF is preferred in $SP$ mode~\cite{Marrero-Santiago:2017b}. Ignition is triggered by laser-induced breakdown allowing a non-intrusive control of the deposit location, duration and strength. Ignition probability maps are constructed using 50 and 30 ignition trials at each deposit location for gaseous cases and the $SP$ case, respectively. This results in a maximum error of the probability of about 7\% and 9\%, respectively.

\begin{table}
\caption{Summary of experimental operating conditions in modes $P$, $NP$ and $SP$. \label{Tbl:expe_cond}}
\centering
\begin{tabular}{| l | c c c |}
\hline
  & P & NP  & SP  \\
\hline
\hline
Central jet $\dot{m}_{Air}$ $(g/s)$ & $0.224$ & - & - \\ 
\hline 
Plenum $\dot{m}_{Air}$ $(g/s)$& $5.37$ & $5.43$ & 8.2 \\ 
\hline
Central jet $\dot{m}_{Fuel}$ $(g/s)$& $0.009$ & $0.234$ & 0.33 \\
\hline
Plenum $\dot{m}_{Fuel}$ $(g/s)$& $0.233$ & - & - \\ 
%\hline
%$S_{w,Expt.}$ &  & $0.76$ & \\
\hline
$\phi_{glob}$ & $0.75$ & $0.75$ & $0.61$ \\
%\hline
%$P_{in}$ (bars) &  & $1.0$ &  \\
\hline
$T_{in}$ (gas) $(K)$& 300 & 300 & 416 \\
\hline
$T_{Fuel}$ (liquid) $(K)$& - & - & 350 \\ 
\hline
\end{tabular}
\end{table}

\subsection{Large Eddy Simulation set-up}
\label{ssec:LESsetup}

All simulations were performed with AVBP, an explicit cell-vertex massively-parallel code solving compressible reacting flows~\cite{Gicquel:2011}. The equations and models used in the present study are standard ones in LES solvers and a full description can be found in the review of Gicquel \emph{et al.}~\cite{Gicquel:2012}. The third order accurate in space and time numerical scheme TTGC~\cite{Colin:2000a} is used. Inlet and outlet boundary conditions are treated according to the NSCBC formulation~\cite{Poinsot:1992}. 
% Félix: ça va perturber les reviewers cette différence de traitement de wall... alors que pour le calcul à froid ça change quasi rien je pense...
% Lucas: Tu penses que ca change quelques chose au spray, à l'évap ? On peux peut-etre rajouter un mot la dessus si besoin.
while non-slipping walls are considered.
%Walls are considered non-slipping in gaseous cases ($P$ and $NP$) and non-slipping iso-thermal ($Tw=387~K$) in the preheated spray mode ($SP$).
% Félix: Les runs reactifs que j'utilise en sections resultats sont dans le futur papier avec Javier, mais pour autant, le setup-up de mon article sympo est valable (sauf LW/TTGC!!) donc je mets cette ref.
% Lucas: C'est bon pour moi !
Turbulent sub-grid stresses are modeled using the SIGMA model~\cite{Nicoud:2011}. In the $SP$ mode, a Lagrangian approach is retained for the dispersed phase description using models for drag, evaporation and injection (FIM-UR model) already presented in a previous study~\cite{Shum-Kivan:2017}. The prescribed droplet size distribution is fitted to experimental data using a Rosin-Rammler distribution
%\begin{align}
%R R \left( d _ { p } \right) &= q \frac { d _ { p } ^ { q - 1 } } { X ^ { q } } \exp \left[ - \left( \frac { d _ { p } ^ { q } } { X } \right) \right] \label{rosin} \\
%X &= \frac { \Gamma ( 1 + 2 / q ) } { \Gamma ( 1 + 3 / q ) } d _ { p } ^ { S M D }. \label{rammler}
%\end{align}  
%\noindent where $\Gamma$ is the standard Gamma function. 
with a spread of the distribution $q = 2.3$ and a mean Sauter diameter is $d _ { p } ^ { S M D } = 31~\mu m$.
%describing the spread of the distribution is set to $q = 2.3$ and the mean Sauter diameter is $d _ { p } ^ { S M D } = 31~\mu m$. %BCFor $P$ and $NP$ cases, chemistry is described with a 2-step mechanism along with the thickened flame approach to model turbulent combustion as detailed in Esclapez \emph{et al.}~\cite{Esclapez:2015}. For the $SP$ case, an ARC chemistry without dedicated combustion model is chosen. Details are provided in Collin-Bastiani \emph{et al.}~\cite{Collin-Bastiani:2018}.

\begin{figure*}[htbp!]
\centering
\includegraphics[width=0.8\textwidth]{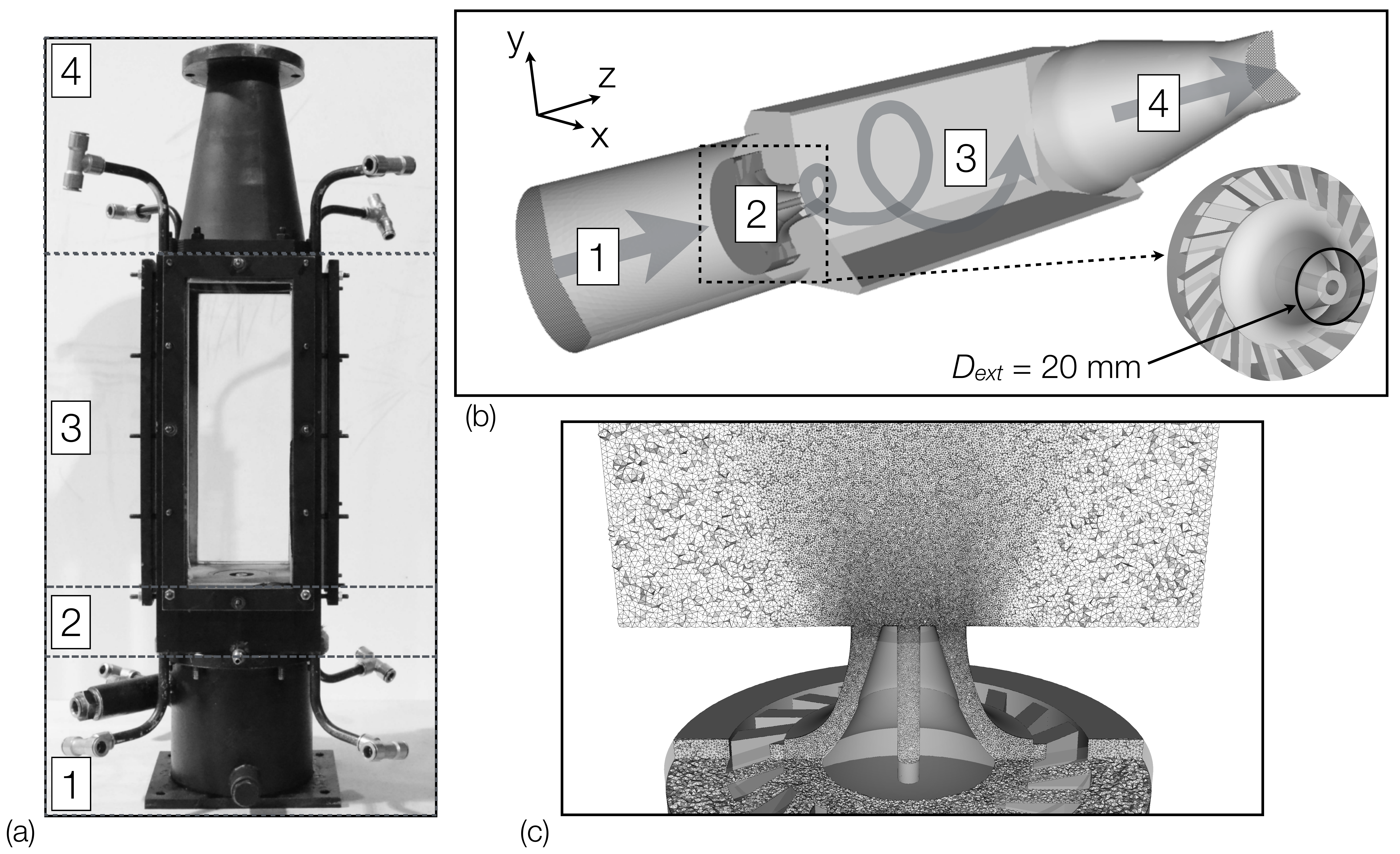}
\caption{(a) Experimental test rig. (b): Numerical geometry and injection system details. Main components are: 1. Plenum, 2. Injection system, 3. Combustion chamber, 4. Convergent exit. (c): Cut through the computational domain showing the mesh refinement near the central gaseous injection ($P$ and $NP$ cases).}
\label{fig:geom}
\end{figure*}

The computational domain includes the four components of the experimental configuration as shown in Fig.~\ref{fig:geom}(b). 
%Félix: Idem, mon maillage est extremement proche de celui de Lucas, donc est-ce que c'est nécessaire d'en parler?
%Lucas: si les tailles de mailles sont similaires dans les zones principales
The domain is discretized into a fully unstructured mesh using 22 million tetrahedral elements shown in Fig.~\ref{fig:geom}(c), with a cell size about 150 $\mu m$ in the swirler and the mixing region and about 800 $\mu m$ in the rest of the combustion chamber. The axial direction is referred to as the $z$-axis, corresponding to the main flow direction, while the $x$-axis and $y$-axis denote the transverse directions. Space dimensions are non-dimensionalized by the injection system exit diameter $D_{ext}$. Flow statistics are collected for over $150~ms$ after reaching the stationary average state.

\subsection{Non reactive LES results}
\label{ssec:LES_NR}

\begin{figure*}[ht!]
\centering
\includegraphics[width=0.85\textwidth]{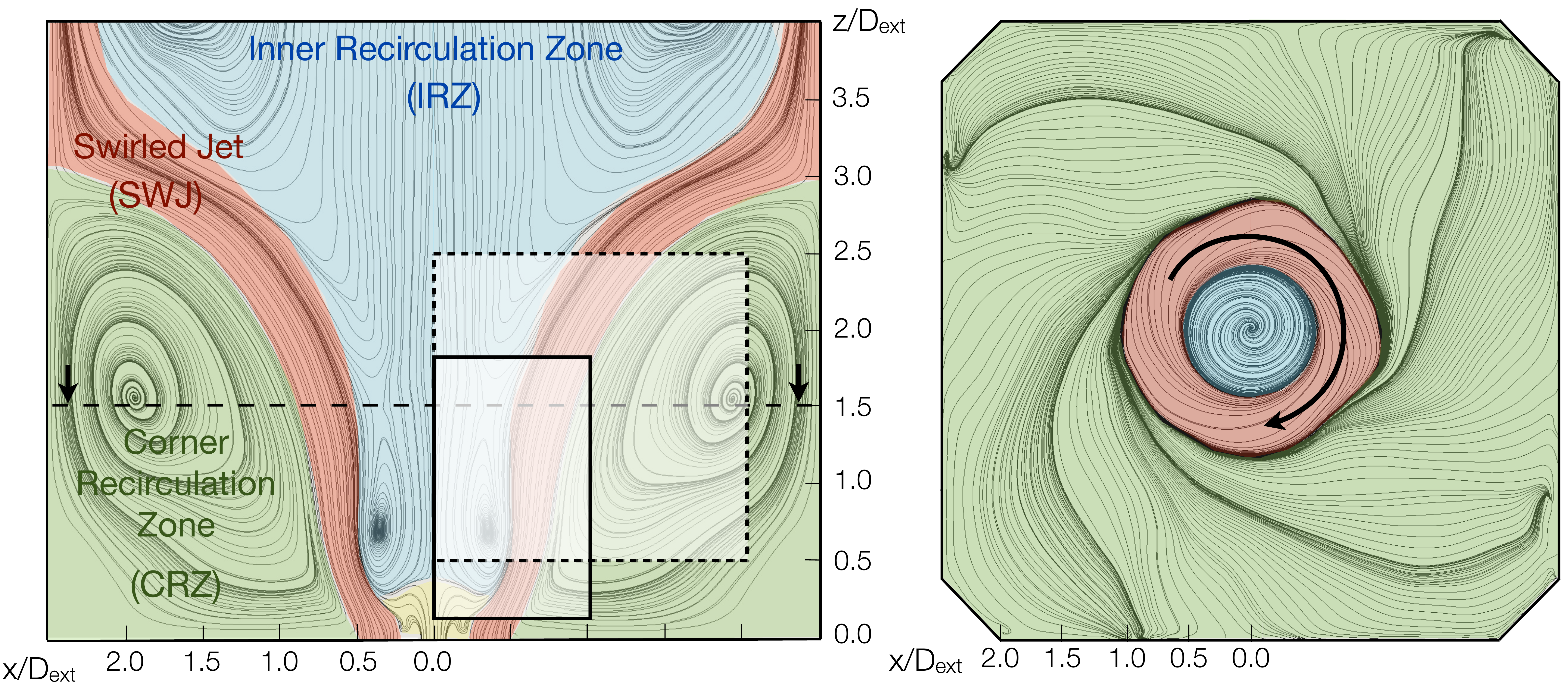}
\caption{$NP$ case, non-reacting flow. Time-averaged pseudo-streamlines in a central $x$-normal plane (left) and $z$-normal plane (right). Swirled Jet (SWJ, red), Inner Recirculation Zone (IRZ, blue) and Corner Recirculation Zone (CRZ, green). Boxes respectively indicate the experimental ignition maps for $P$, $NP$ (plain) and $SP$ (dashed) cases.}
\label{fig:stream}
\end{figure*}

The flow pattern shown in Fig.~\ref{fig:stream} is typical of highly swirled configurations: the Swirled Jet (SWJ) issued from the injection system generates a reverse flow along the central axis referred to as Inner Recirculation Zone (IRZ). The IRZ closes downstream at $z/D_{ext}=10$ due to the presence of the convergent exhaust. Because of the confined environment, the SWJ also induces recirculation on its outer side, referred to as Corner Recirculation Zones (CRZ), closed at $z/D_{ext}=3$ in the gaseous cases and at $z/D_{ext}=2.5$ in the $SP$ case. 
The gaseous flow exiting the central injection in $P$ and $NP$ cases meets the back flow of the IRZ at $z/D_{ext}=0.4$, generating a zero axial velocity stagnation point. In the $SP$ case, the $n$-heptane injection momentum leads to a stagnation point almost at the injector surface. Finally, strong shear layers develop between the SWJ and both IRZ and CRZ. Note that the appearance of vortex breakdown and the formation of the IRZ occurs as the swirl number exceed a critical value ($S_{w,crit} = 0.707 $~\cite{Billant:1998}). In the $NP$ case, the swirl number has been measured experimentally $S_{w,Exp.} = 0.76$ and a very close value $S_{w,LES} = 0.78$ has been computed from the LES results.

% in the $NP$ case, the swirl number $S_w$ evaluated in LES, $S_{w,LES} = 0.78$, is very close to the experimental value $S_{w,Exp.} = 0.76$, above the theoretical critical value for vortex breakdown and IRZ formation .

Detailed comparison of the non-reacting LES prediction against experiment for the $P$ case has been reported in a previous publication~\cite{Barre:2013}. Similar comparison is presented in~\ref{app:valid_cold} for the $NP$ and $SP$ cases. All show a very good agreement and authorize the development of the ignition model on the basis of LES results.

\begin{figure*}[ht!]
\centering
\includegraphics[width=0.7\textwidth]{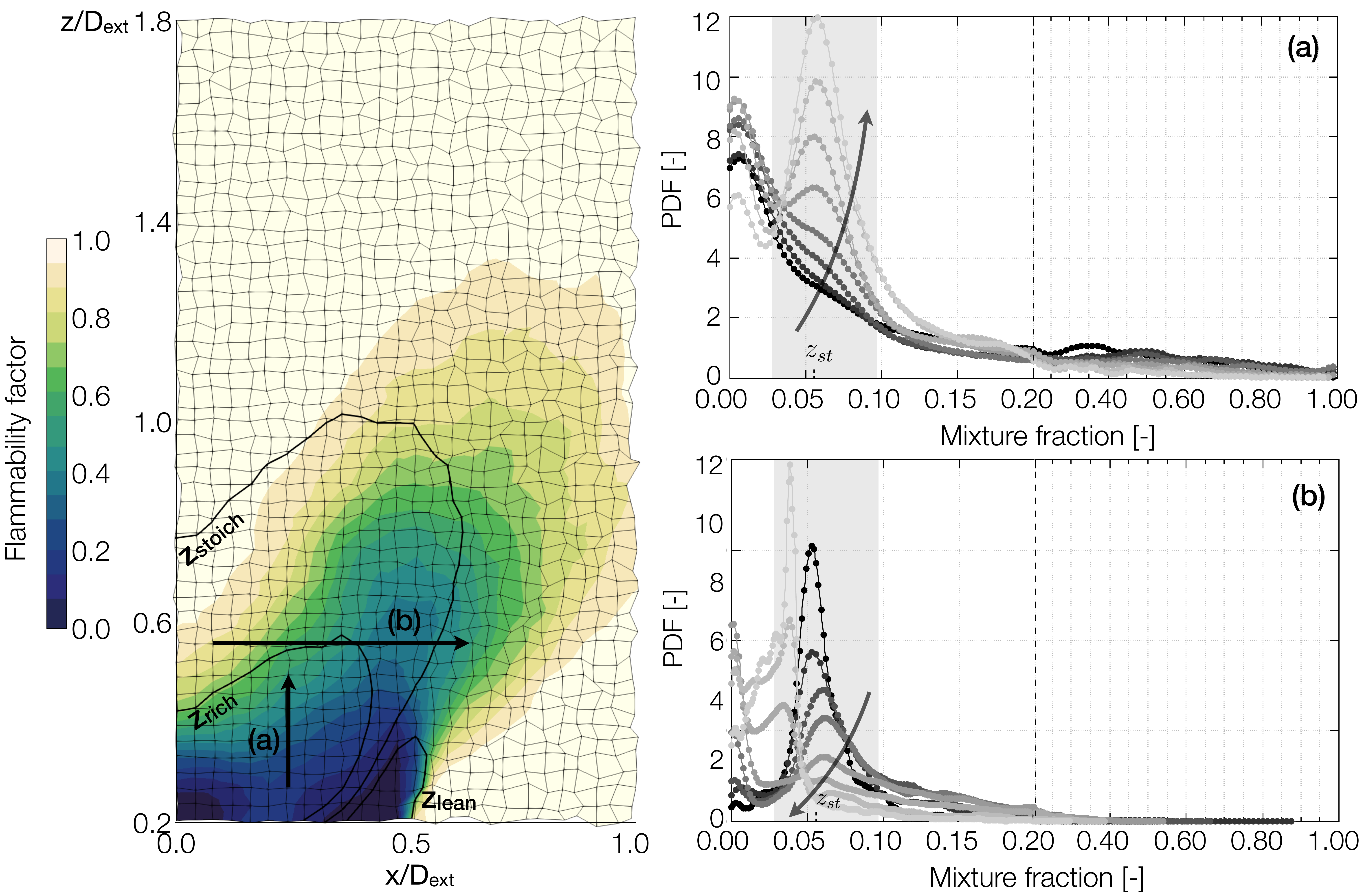}
\caption{$NP$ case. Mean flammability factor field in a central $x$-normal plane with $Z$ iso-lines (left) and $P(Z)$ along arrows (a) (top) and (b) (bottom) in the mixing region (right). The grey area highlights the flammable mixture interval.}
\label{fig:Mixing}
\end{figure*}

% Félix: J'ai mis pas mal de temps à capter que les flèches (a) et (b) correspondent aux 2 graphes à droite... Ça vaut peut etre le coup de reformuler pour que ce soit plus évident?
%Félix: Remettre les dimensions en x/Dext et en z/Dext ? à moins que ça corresponde pile à la box de la fig 2? Mais dans ce cas, il faut l'indiquer je pense
% Yep ! J'ai rajouter les coordonées
%In order to further characterize the mixing process in the non-premixed case at the vicinity of the central jet,
A focus is now made on mixing, which is critical for ignition in both $NP$ and $SP$ cases. Fig.~\ref{fig:Mixing}(left) shows the mean flammability factor for case $NP$:
\begin{equation}
F_f = \int_{Z_{lean}}^{Z_{rich}} P(Z)\;dZ
\label{eq:Ff}
\end{equation}
\noindent where $P(Z)$ is the probability density function (PDF) of the mixture fraction $Z$ (using the definition of Bilger \cite{Bilger:1989}) and $Z_{lean}$ and $Z_{rich}$ are the lower and upper flammability limits, respectively. Since the overall equivalence ratio is flammable, $F_f$ is unity in most of the combustion chamber where all the species are well mixed, and reaches 0 only close to the methane and air inlets. Intermediate values of $F_f$ are found in the wake of the air swirled jet, between the rich injection and the pure air. The IRZ is mostly filled with premixed flammable mixture. The mixture fraction PDF extracted along the arrows (a) and (b) of Fig.~\ref{fig:Mixing}(left) and displayed in Fig.~\ref{fig:Mixing}(right) show the variety of $P(Z)$ and the strong inhomogeneity in these zones.

\begin{figure}[ht!]
\centering
\includegraphics[width=0.48\textwidth]{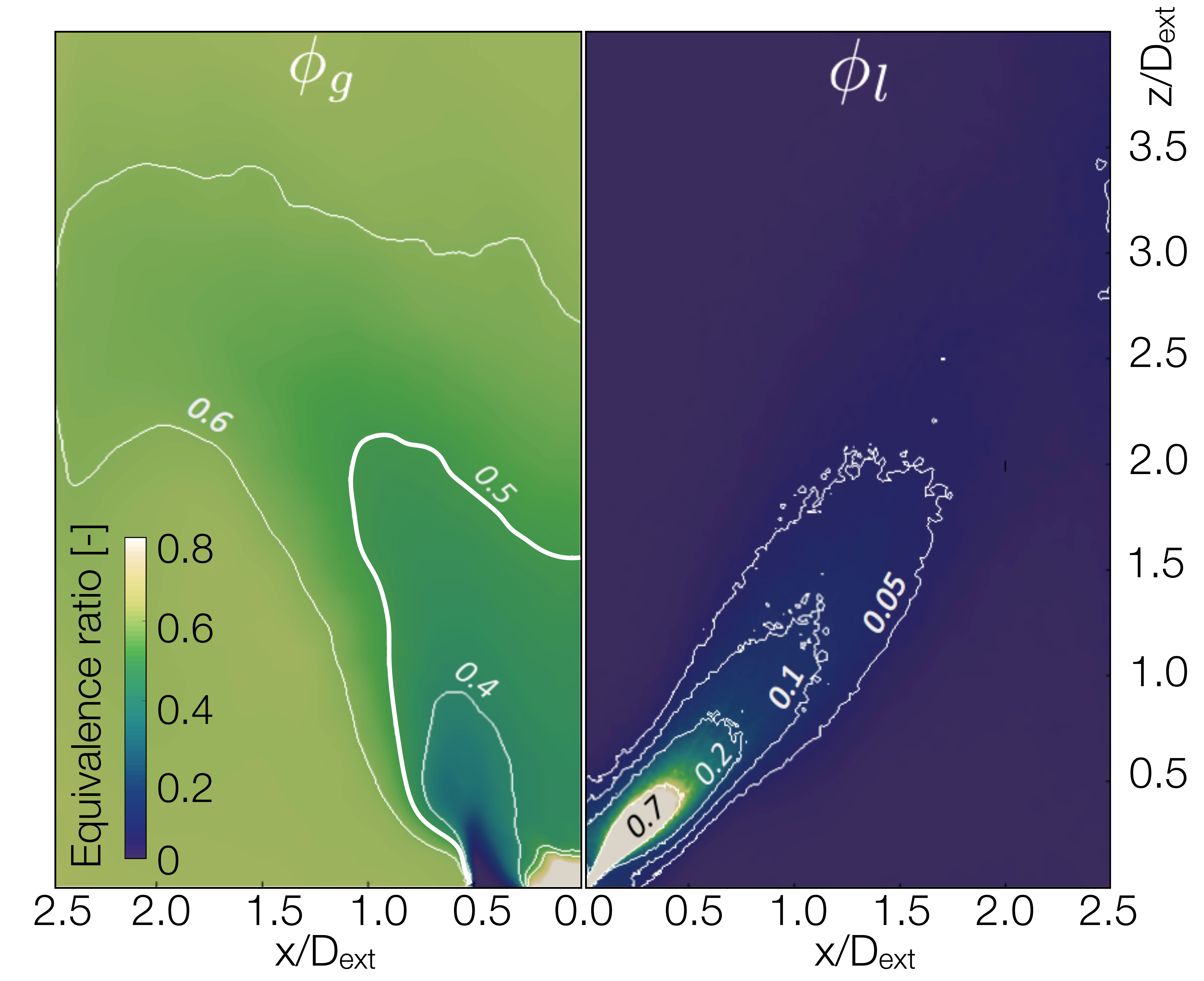}
\caption{$SP$ case. Maps of the cold flow gaseous equivalence ratio $\phi_g$ (left) and liquid equivalence ratio $\phi_l$ (right). The highlighted $\phi_g = 0.5$ corresponds to the lean flammability limit of \emph{n}-heptane which also marks the transition between weakly evaporation-controlled flames ($\phi_g > 0.5$) and evaporation-controlled flames (Section \ref{sssec:mixture}).}
\label{fig:evap_mix_spray}
\end{figure}

In the $SP$ case, evaporation and mixing effects reflect on the gaseous and liquid equivalence ratio maps $\phi_g$ and $\phi_l$, shown in Fig.~\ref{fig:evap_mix_spray}. Due to the preheated conditions, droplets evaporate quickly leading to $\phi_l > 1$ in the spray jet zone for $z/D_{ext}< 1$. Almost no droplets are found in the upper part of the spray zone and even less in the IRZ and CRZ. The entire CRZ is characterized by a very homogeneous gaseous equivalence ratio close to the global value $\phi_{glob} = 0.61$, whereas the IRZ is leaner ($\phi_g$ < 0.5), close to the lean flammability limit.  

\section{The MIST Model}
\label{sec_model}

The prediction of the ignition probability is classically based on the combination of kernel motion statistics with local flow properties. However in contrast with previous methodologies~\cite{Neophytou:2012, Linassier:2013, Eyssartier:2013}, here the flame kernel trajectory statistics are built from the non-reacting flow statistics. The objective of MIST is to predict the probability of creating a large enough flame kernel, that can subsequently stabilize on the injector. Capturing the flame stabilization process itself is not in the scope of MIST since LES has shown that the kernel expansion can significantly modify the instantaneous velocity field in the combustor \cite{Jones:2010,Barre:2014}, rendering inaccurate the cold flow statistics upon which MIST is based. Although failure to stabilize the flame after the kernel occupies a significant portion of the combustion chamber has been observed experimentally \cite{Read:2008}, we believe this mode of failure marginally affects the overall ignition probability compared to the critical stage of creating an expanding flame kernel. In the present experimental test case, such failure mode was not observed. Additionally, ignition stochasticity mostly occurs in the first instants of ignition, when local turbulence and mixing along the kernel trajectory completely control the flame kernel survival, whereas at later time local turbulence and mixture composition only affect the ability of flame fronts to propagate locally.

The model can be decomposed in four steps:
\begin{enumerate}
\item Extract from a non-reacting flow solution the mean and rms of the velocity ($\overline{\boldsymbol{u}}$, $\boldsymbol{u'}$) and mixture fraction ($\overline{Z}$, $Z'$). Liquid volume fraction moments ($\overline{\alpha_l}$, $\alpha_l'$), mean droplet diameter ($\overline{d_l}$), and mean droplet velocity ($\overline{\boldsymbol{u_l}}$) are also required for the $SP$ case. If LES is used, statistics are obtained from time-averaging.
\item Use the spark characteristics to evaluate the kernel initial size and the time required for cooling from the sparking temperature to the burnt gas temperature. This step is performed in 0D assuming that the kernel temperature evolution is dictated by the balance between combustion heat release and turbulent dissipation.
\item Compute quenching criteria from the non-reacting flow statistics.
\item Starting from the initial kernel defined in step 2, compute the temporal evolution of kernel motion statistics. This is based on the evolution of the kernel probability of presence $P_{pres}$ constructed from flow statistics obtained in step 1 and the quenching criteria computed in step 3. In this step the kernel size evolution is also computed to determine when it has grown sufficiently to ensure a successful ignition. 
\end{enumerate}

Note that step 1 may be performed with any approach able to give flow statistics, either numerically or with measurements. A flowchart summarizing the main MIST steps described hereafter is provided in ~\ref{app:MISTflowchart}.

\subsection{Step 2: Initial kernel}
\label{ssec:initkernel}
Following the spark discharge, the transition between the hot plasma and a self-sustained flame kernel occurs at temperatures largely above the burnt gas temperature~\cite{Maly:1978}. A detailed description of this transition requires to take into account complex physico-chemical interactions and is out of the scope of the present model. Here the initial kernel development is split in two phases: the kernel growth is first sustained by the high temperature associated with the energy deposit, then it is driven by combustion. During the first phase, the kernel can survive a non-flammable mixture or strong turbulence. This has been observed experimentally in typical gas turbine configuration \cite{Mastorakos:2009} and more recently further studied in a stratified turbulent flow configurations \cite{Sforzo:2015, Sforzo:2017}, where the spark igniter is located in a non-flammable region and the kernel transition from this adverse location to a flammable region is studied. A data-driven model to predict the behavior of the flame kernel during that transition was proposed, highlighting the importance of cold gas entrainment in the kernel \cite{Sforzo:2017}. Such process is not accounted for in the present modeling approach, but could be investigated to adapt the present model to various types of ignition system. The simple model described hereafter aims at evaluating the time required for the kernel to cool down to the burnt gas temperature which will be used in Step 4 to apply extinction criteria.
Given the amount of deposited energy $\varepsilon_i$ and the deposit volume $V_s$, the initial kernel temperature $T_{k}^0$ is given by (assuming no reaction during the short deposition duration):
\begin{equation}
T_{k}^0 = T^0 + \frac{1}{\rho C_p}\frac{\varepsilon_i}{V_s}
\end{equation}
where $T^0$ is the initial gas temperature, and $\rho$ and $C_p$ are respectively the initial gas density and specific heat. In practice, the computation described hereafter is performed using standard thermodynamics, which are not suited for high-temperature plasma. The maximum temperature is then limited to 5000 K, from which it is possible to evaluate the initial kernel radius assuming that the spark deposit is Gaussian in space (classically used in many DNS and LES of ignition events, see \cite{Lacaze:2009b} for more details). The spark energy used in MIST matches standard value used in previous LES \cite{Esclapez:2015,Collin-Bastiani:2018}: 30 mJ in the $P$ and $NP$ cases, and 25 mJ in the $SP$ case.

\begin{figure}[ht!]
\centering
\includegraphics[width=0.48\textwidth]{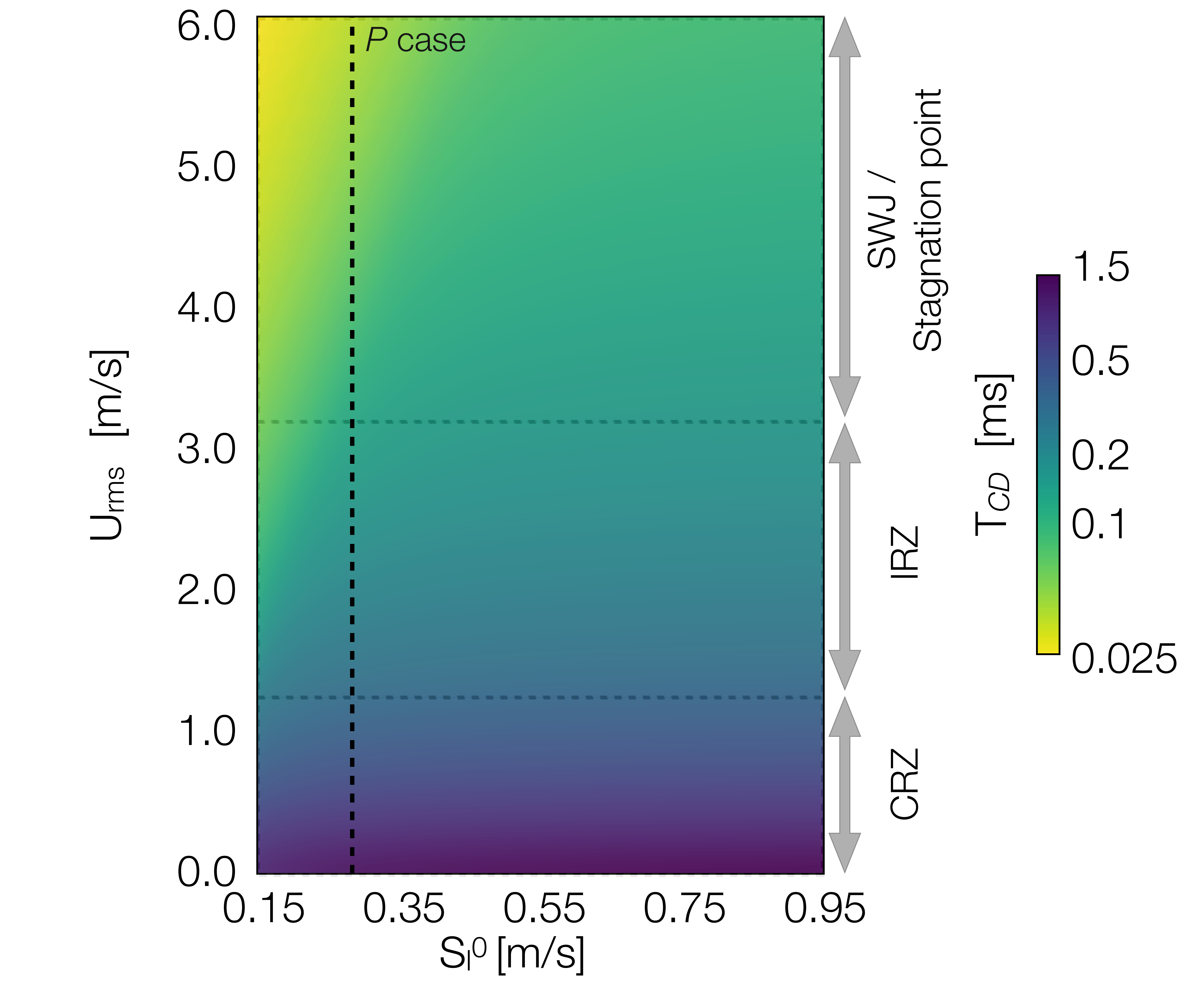}
\caption{Kernel cooling time $t_{CD}$ map as function of $S_L^0$ and $u'$ for an initial methane/air kernel with a spark energy of 30 mJ. The vertical dashed line corresponds to the $P$ case laminar flame speed. The vertical arrows indicate typical range of $u'$ in distinct areas of the swirled flow (see Fig.~\ref{fig:stream}).}
\label{fig:step2}
\end{figure}

The kernel temperature $T_k$ then evolves following a 0-dimensional equation:
\begin{equation}
\frac{dT_k}{dt} = \dot{\omega}_T(\overline{Z}_{flam}) + \frac{D_{th}}{r_k^2}(T^0 - T_k)
\label{eq_kernel_balance}
\end{equation}
The combustion heat release rate $\dot{\omega}_T$ is evaluated at the mean flammable mixture fraction $\overline{Z}_{flam}$ in the sparking zone using the laminar flame expression: 
\begin{equation}
\dot{\omega}_T(\overline{Z}_{flam}) = \frac{Y_F(\overline{Z}_{flam}) \mathcal{Q}_r S_L^0(\overline{Z}_{flam})} {C_p \delta_L^0(\overline{Z}_{flam})} \label{eq:source_combu}
\end{equation}
with
\begin{equation}
\overline{Z}_{flam} = \frac{\int_{Z_{lean}}^{Z_{rich}} Z P(Z)\;dZ}{F_f} \label{zflam}
\end{equation}
In Eq.~\ref{eq:source_combu}, $\mathcal{Q}_r$ is the heat of combustion, and $S_L^0$ and $\delta_L^0$ are the laminar flame speed and thickness. The diffusive heat loss $D_{th}$ is computed with the sum of laminar and turbulent thermal diffusivities, the latter given by~\cite{Akindele:1982}:
\begin{align}
D_{th,turb} =  0.44 u' l_t \left( 1 - \exp \left( -\frac{u' t}{0.44 l_t} \right) \right)
\end{align}
where $l_t$ is the integral turbulent scale. The turbulent diffusivity progressively increases with time $t$ from 0 to its fully developed value, in order to reflect that, with time, the kernel interacts with turbulent eddies of increasing size~\cite{Akindele:1982}. Finally, the kernel growth is simply calculated using the laminar flame speed~\cite{Boudier:1992}:
\begin{equation}
\frac{dr_k}{dt} = \frac{T_k}{T^0}S_L^0(\overline{Z}_{flam}) \label{eq:drk/dt}
\end{equation}
%The flow characteristics at the spark location are used as inputs for the temperature balance equation: the kernel displacement during this short initial phase is neglected. 
Resolving Eq.~\ref{eq_kernel_balance} with the flow properties at the spark location leads to the kernel cooling time $t_{CD}$. For two-phase ignition, $S_L^0$ is simply replaced by $S_L^{tp}$ \cite{Rochette:2018} in Eqs.~\ref{eq:source_combu} and~\ref{eq:drk/dt}. To illustrate the outcome of this process, Fig.~\ref{fig:step2} shows $t_{CD}$ as function of $S_L^0$ and $u'$ for a spark energy of 30 mJ and a constant integral length scale of 1 cm. The gas properties used to obtain these results correspond to that of methane/air mixtures, but the range of laminar flame speed has been extended to provide a more complete picture. The range of $u'$ was extracted from the non-reacting LES: the low velocity CRZ are characterized by low levels of turbulence, where $t_{CD}$ can reach around 1 ms, whereas in the highly turbulent shear layer of the SWJ or at the vicinity of the stagnation point, high turbulence level induces a rapid drop of the initial kernel temperature corresponding to a cooling time of the order 10$\sim$100 $\mu$s.

\subsection{Step 3 : Quenching criteria}
\label{ssec_indics}

Following previous studies, two major mechanisms leading to kernel quenching are considered: mixing ~\cite{Birch:1977} and flame stretching~\cite{Wilson:1999,Neophytou:2012}.

\subsubsection{Mixture composition}
\label{sssec:mixture}

% Félix: j'introduis des paragraphes car sinon c'est beaucoups d'infos d'un coup. C'est ici que je parle en détail de l'aspect TPF...
\paragraph{Gaseous cases} %\mbox{}\\

Several ignition studies in non-pre\-mixed flow in the literature clearly point out the fact that the flammability factor $F_f$ is a critical parameter~\cite{Birch:1977, Ahmed:2007a, Neophytou:2012}, closely related to the probability of creating a sustainable flame kernel. As performed in experimental studies~\cite{Birch:1977,Birch:1981}, time-averaged statistics $\overline{Z}$, $Z'$ obtained here from the non-reacting LES are used to construct the flammability factor. It requires however to assume a shape for the probability density function $P(Z)$. In free jets, the combination of the Gaussian and Dirac functions provides a fairly good estimate of $F_f$~\cite{Schefer:2011}. For more complex cases such as swirled flows, the large variety of mixture fraction distributions (see Fig.~\ref{fig:Mixing}) is better represented by a combination of the log-normal and $\beta$-distributions:
\begin{equation}
F_{f,model} = \gamma F_{f,\beta} + (1 - \gamma) F_{f,logN}
\label{eq:Ffmodel}
\end{equation}
where
\begin{align}
F_{f,logN} = \frac{1}{2} \Big[ &\text{erf}\left( \frac{\ln(Z_{rich})-\overline{Z}}{\sqrt{2}\;Z'^2}\right)  \nonumber \\
- &\text{erf}\left( \frac{\ln(Z_{lean})-\overline{Z}}{\sqrt{2}\;Z'^2}\right) \Big]
\end{align}
is the log-normal cumulative distribution function, and 
\begin{equation}
F_{f,\beta} = \frac{B_{Z_{rich}}(\alpha,\beta)}{B(\alpha,\beta)} - \frac{B_{Z_{lean}}(\alpha,\beta)}{B(\alpha,\beta)}
\end{equation}
where $B_{z}(\alpha,\beta)$ is the incomplete $\beta$ function of parameters $\alpha$ and $\beta$ given by:
\begin{equation}
\alpha = \overline{Z}\left( \frac{\overline{Z}(1-\overline{Z})}{Z'}-1 \right) ;
\beta = (1- \overline{Z})\left( \frac{\overline{Z}(1-\overline{Z})}{Z'}-1 \right)
\end{equation}

The blending factor $\gamma$ is designed so as to make a transition from regions of low mixture fraction fluctuations, where $F_{f,logN}$ is preferred, to regions of high fluctuations where $F_{f,\beta}$ is applied:
\begin{equation}
\gamma = 0.5 \left(1 + \text{tanh}\left( \frac{Z' - \overline{Z}}{Z_{glob}}\right) \right)
\end{equation}

Note that $P(Z)$ may be directly extracted from the non-reacting LES. %the derivation and validation of a method to accurately reconstruct $P(Z)$ for a wide range of flow topology is out of the scope of the present work. 
The above method is mostly presented for completeness of the model, and to highlight the importance of including the impact of recirculating gas in the local mixture composition, which was not accounted for in previous studies. The mixture fraction PDF allows to also compute $\overline{Z}_{flam}$ (Eq.~\ref{zflam}), which is a second important quantity for ignition. The accuracy of the predicted values of $F_f$ and $\overline{Z}_{flam}$ in $NP$ case is demonstrated by comparison to the actual values obtained from LES in \ref{app:mixt_stat}.

%\begin{equation}
%\overline{Z}_{flam} = \frac{\int_{Z_{lean}}^{Z_{rich}} Z P(Z)\;dZ}{F_f} \label{zflam}
%\end{equation}
%Taking $P(Z)$ as the composite PDF introduced previously, $\overline{z}_{flam}$ is found:
%\begin{align}
%\overline{z}_{flam}  = & \; \left( \gamma \int_{z_{lean}}^{z_{rich}} \frac{1}{B(\alpha,\beta)} z^{\alpha} (1-z)^{\beta-1} \;dz \right. \nonumber \\
%& \left.+ (1-\gamma) \int_{z_{lean}}^{z_{rich}} \frac{1}{\sigma\sqrt{2\pi}}\exp^{-\frac{(\ln(z)-\mu)^2}{2\sigma^2}} \;dz \right) / F_{f,model}
%\label{eq:MeanZflam}
%\end{align}

\paragraph{Spray cases}% \mbox{}\\

In addition to the directly available gaseous fuel, $F_f$ must take into account the evaporating liquid fuel. The characteristic evaporation time:
\begin{equation}
\tau _ { e v } = \frac { \rho _ { l } d _ { p} ^ { 2 } } { 8 \rho _ { g } D _ { F } \ln \left( 1 + B _ { M } \right) } 
 \label{eq:tauev}
\end{equation}
is compared to the characteristic combustion time $\tau_{c}(\phi) \approx \delta_l^0(\phi) / S_L^0(\phi)$. In Eq.~\ref{eq:tauev}, $\rho_l$ and $\rho_g$ are the liquid and gaseous densities, $D_F$ is the fuel diffusivity and $B_M$ is the Spalding mass transfer number. Both Eqs.~\ref{eq:Ffmodel} and~\ref{zflam} still hold, with however a modified $F_f$ as described below.

Depending on the ratio of the fresh gas equivalence ratio $\phi_g$ to the lean flammability limit $\phi_{lean}$, two archetypes of two-phase kernels are distinguished: weakly evaporation-controlled flames and evaporation-controlled flames:

\begin{itemize}
\item A weakly evaporation-controlled flame corresponds to $\phi_g > \phi_{lean}$, or where liquid fuel evaporates very promptly:
\begin{equation}  
U^* \frac{\tau_{ev}}{\tau_c} < 1 \label{Ustar}
\end{equation}
with $U^* = u_l/u_g$  the relative velocity between fuel droplets and the carrier phase. Such a flame is very close to a purely gaseous flame and $F_f$ is estimated as in the gaseous case with Eq.~\ref{eq:Ffmodel} where $\overline{Z} = \overline{Z_{eff}}$ includes the evaporated fuel consumed in the flame of thickness very close to $\delta_l^0$~\cite{Rochette:2018}:
\begin{equation}
%\overline{Z_{eff }} = \overline{Z_g} + \left( \frac { \delta _ { l } ^ { 0 } } { \max \left( \delta _ { e v } , \delta _ { l } ^ { 0 } \right) } \right) ^ { 2 / 3 } \overline{Z _ { l }} = \overline{z _ { g }} +  \Gamma \, \overline{z _ { l }}  \label{zeff}
%\overline{Z_{eff }} = \overline{Z_g} + \left( \frac { \delta _ { l } ^ { 0 } } { \max \left( \delta _ { e v } , \delta _ { l } ^ { 0 } \right) } \right) ^ { 2 / 3 } \overline{Z _ { l }} \label{zeff}
\overline{Z_{eff }} =  \overline{Z _ { g }} +  \Gamma \, \overline{Z _ { l }}  \label{zeff}
\end{equation}
with $\overline{Z _ { l }}$, $\overline{Z_ { g }}$ the mean liquid and gaseous mixture fractions, and:
\begin{equation}
\Gamma =  \left( \frac { \delta _ { l } ^ { 0 } } { \max \left( \delta _ { e v } , \delta _ { l } ^ { 0 } \right) } \right) ^ { 2 / 3 } 
\end{equation}
where  $\delta _ { e v } = u_l * \tau_{ev}$ is the evaporation length. The fluctuating mixture fraction $Z_{eff}'$ originates from turbulent mixing and spray local evaporation. It is assumed here that cold flow evaporation is negligible compared to evaporation in the flame, so that $Z_{eff}'$ may be evaluated as:
\begin{equation}
Z_{eff}' = \underbrace {Z_g'}_{\substack{\text{turbulent} \\ \text{mixing}}} + \underbrace {\Gamma \frac{\rho_l}{\rho_g} \alpha_l'}_{\substack{\text{evaporation} \\ \text{in the flame}}} \label{zefffluctu}
\end{equation}
where $Z_g'$ and $\alpha_l'$ are again obtained from the non-reacting flow statistics.

%Finally, the conditional mean flammable mixture fraction $\overline{z}_{flam}$ is calculated as in gaseous flows using Eq.~\ref{zflam}.

\item An evaporation-controlled flame corresponds to $\phi_g < \phi_{lean}$. In that case evaporation is the limiting process in the flame:
\begin{equation}  
U^* \frac{\tau_{ev}}{\tau_c} > 1 \label{weak_case}
\end{equation}
As a consequence the consumption rate decreases compared to the previous case, and the liquid fuel is burnt as soon as it is evaporated, leading to:
\begin{align}
\overline{Z _ { e f f }} &= \overline{Z _ { l }} + \overline{Z _ { g }}  \label{zeff_evapcontrolled}, \\
Z_{eff}' &= Z_g' + \frac{\rho_l}{\rho_g} \alpha_l'. \label{zeff_fluctu_evapcontrolled}
\end{align}

Note that in the present configuration, the evaporation-controlled formulation is only used near the spray injection, where the amount of fuel pre-vaporized is below the flammability limit of \emph{n}-heptane (see Fig.~\ref{fig:evap_mix_spray}). It is expected to become more significant in realistic configuration where the incoming air temperature is lower and the volatility of the fuel might be lower. In particular, altitude relight conditions are characterized by low temperature at which very little evaporation occurs prior to ignition and for which the evaporation-controlled formulation is especially adapted.
\end{itemize}

\subsubsection{Flame stretch}

Flame / turbulence interaction may be responsible of significant quenching due to fragmentation of the flame kernel. Following the previous works of~\cite{Wilson:1999,Neophytou:2012}, a criterion based on the Karlovitz number is used. The estimation of $Ka$ is taken from ~\cite{Abdel-Gayed:1985}:
\begin{equation}
Ka = 0.157\left( \nu \varepsilon \right) ^{1/2} \frac{1}{{S_L^0}^2}
\label{eq:Ka_abdel}
\end{equation}
where $\varepsilon$ is the turbulent dissipation, $\nu$ is the kinematic viscosity. For $SP$ cases, $S_L^0$ is replaced by the two-phase laminar flame speed $S_L^{tp}$ proposed in~\cite{Rochette:2018}. For weakly evaporation controlled flames, $S_L^{tp}\sim S_L^0(\overline{Z}_{flam})$. For evaporation controlled flames $S_L^{tp}$ is much smaller that $S_L^0$ and can be estimated by replacing $\tau_c$ by $\tau_{ev}$:
\begin{equation}  
S _ { L } ^ { t p } = \frac { \delta ^ { 0* } _ { l } } { \tau _ { e v } } \label{slevap},
\end{equation}
where $\delta ^ { 0* } _ { l }$ is the flame thickness at the equivalence ratio $\phi^*=min(\phi_{tot},1)$, with $\phi_{tot}=\phi_g + \phi_l$ the total equivalence ratio.

The turbulent dissipation $\varepsilon$ may be directly extracted from LES or reconstructed from $\overline{u}$ and $u'$ fields. In the latter case series of instantaneous velocity fields, and their dissipation rate tensor, may be reconstructed assuming a Gaussian distribution. Taking the average over 20-50 reconstructed velocity fields is generally sufficient to ensure a statistically converged value of  $\varepsilon$.

%However, evaluating $\varepsilon$ from the mean velocity field is inadequate. Therefore, under the hypothesis of Gaussian instantaneous velocity distribution around $\overline{u_i}$ with a variance $u_i'$, many instantaneous velocity fields $\Tilde{u}$ are reconstructed based on the non-reacting flow time-averaged data.
%$\varepsilon$ is then the average of all individual dissipation rate tensor $\Tilde{\varepsilon}$ associated to each $\Tilde{u}$ field. 20-50 reconstructed $\Tilde{u}$ fields are generally sufficient to ensure a converged $\varepsilon$ tensor. %The validity of the hypothesis of Gaussian shaped velocity is provided in~\ref{app:velo}. 
%%%%%%%%%%%

Quenching occurs when the Karlovitz number is above a critical value $Ka_c$. Different values of $Ka_c$ are proposed in the literature.  
%A critical value $Ka_c$ must then be determined, above which the flame is supposed to quench.
A value of $Ka_c = 1.5$ is reported in~\cite{Abdel-Gayed:1985, Neophytou:2012} for premixed flames. In \cite{Cordier:2013}, the best agreement of the ignition model with experimental data leads to $Ka_c = 4.5$. This latter value is retained in the present work as it resulted in best overall agreement between MIST and the set of experiment data. Further tuning of this parameter could be required in configurations having flow features not included in the present configuration.

%As this value appears more consistent with our $Ka$ field, it is retained. The extinction criteria based on the Karlovitz number is then defined:
%\begin{equation}
%C_{Ka} = \left\{ \begin{array}{rl} 1 & \text{if} \; Ka<Ka_c \\
%0 & \text{otherwise} \end{array} \right.
%\end{equation}

\subsection{Step 4 : Kernel trajectories}
\label{ssec_ppres}

In previous ignition model \cite{Neophytou:2012}, statistics of kernel trajectories were computed using a Monte-Carlo approach, calculating numerous ignition events and kernel trajectories. In contrast, the PDF of presence $p(\boldsymbol{x},r,t)$ of kernels of size $r$ at the location $\boldsymbol{x}$ and time $t$, is here directly obtained from the non-reacting flow statistics. To do so, four assumptions are made:
\begin{itemize}
\item the velocity components follow a Gaussian distribution,
\item kernel trajectory statistics follow a Markov process,
\item velocity statistics of the non-reacting flow remain valid during the first instants of ignition (before thermal expansion appears),
\item the flame speed is low compared to the flow velocity.
\end{itemize}

\begin{figure*}[ht!]
\centering
\includegraphics[width=0.8\textwidth]{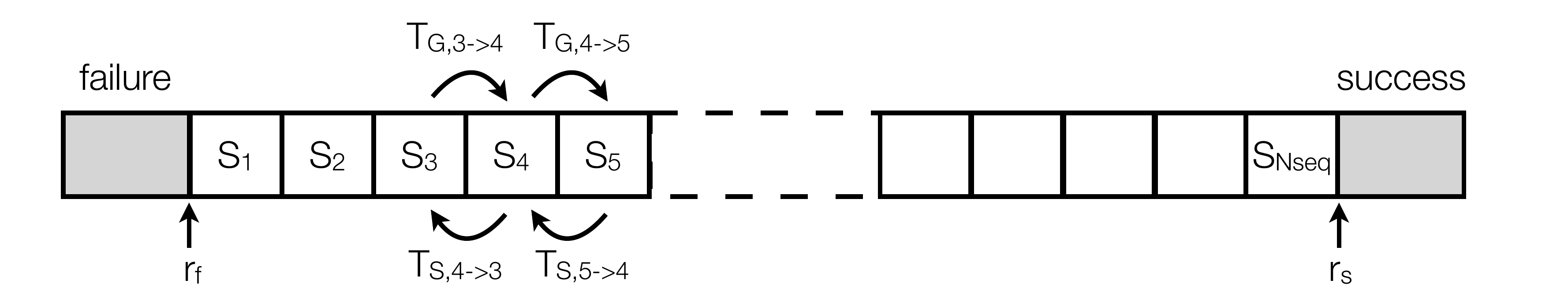}
\caption{Breakdown of the kernel size space into sections with transfer  rates between consecutive sections.}
\label{fig:radius_sections}
\end{figure*}

As often made for particle statistics, the PDF $p(\boldsymbol{x},r,t)$  is discretized in $r$-space using $N_{sec}$ sections $S_i$ as depicted in Fig.~\ref{fig:radius_sections}. In each section $i$, $p(\boldsymbol{x},r,t)=p_{i}(\boldsymbol{x},t)$ is constant. 

From the second assumption,
 the position $\boldsymbol{x}(t)$ of a kernel follows the Langevin stochastic differential equation (SDE)~\cite{Boughton:1987}:
\begin{equation}
\frac{\text{d} \boldsymbol{x}(t)}{\text{d} t} = \mu(\boldsymbol{x}) + \sigma(\boldsymbol{x})\eta(t),
\label{eq:Langevin}
\end{equation}
where the initial kernel position $\boldsymbol{x}(t=0) = \boldsymbol{x}_0$ is the spark position. The function $\mu(\boldsymbol{x})$ corresponds to the deterministic (mean) motion while the second term introduces the turbulence effect. $\eta(t)$ is a white noise (stationary, Gaussian random process with zero mean and delta-Dirac autocorrelation). 
the temporal evolution of $p_i(\boldsymbol{x},t)$ is governed by the Fokker-Planck equation~\cite{Gardiner:2009}:
\begin{align}
\frac{\partial p_i(\boldsymbol{x},t)}{\partial t} = -& \frac{\partial }{\partial \boldsymbol{x}} (\mu(\boldsymbol{x}) p_i(\boldsymbol{x},t)) \nonumber \\
+& \frac{1}{2} \frac{\partial^2 }{\partial \boldsymbol{x}^2}  \left( D_p (\boldsymbol{x}) p_i(\boldsymbol{x},t) \right)  \nonumber \\
+& \dot{Q}_i 
\label{eq:FP}
\end{align}
%where the first term on the RHS represents the mean drift of $p(\boldsymbol{x},r_k,t)$ by the underlying flow, while the second term represents turbulent diffusion \cite{Boughton:1987, Gardiner:2009}.
%As pointed out previously, the Fokker-Planck equation for the Markov displacement process is related to advection-diffusion transport equation of a scalar concentration.
The function $\mu(\boldsymbol{x})$ corresponds to the deterministic (mean) motion while the $D_p (\boldsymbol{x})$ introduces turbulence diffusion. These parameters are related to the flow statistics by:
%assumed to be stationary and can be computed from the velocity statistics by rewriting Eq.~\ref{eq:Langevin} as follows:
%\begin{equation}
%\text{d} \boldsymbol{x}_k(t) = \mu(\boldsymbol{x}_k)\text{d}t + \sigma(\boldsymbol{x}_k)\text{d}W(t), \; \boldsymbol{x}_k(t=0) = \boldsymbol{x}_0
%\label{eq:Langevin_v2}
%\end{equation}
%where $\text{d}W(t)$ is an increment of the Wiener process corresponding to the white noise $\eta(t)$. Using the fact that $<\text{d}W(t)> = 0$ and $<\text{d}W(t)^2> = \text{d}t$ (where $< >$ denotes an ensemble average over many realizations of the process):
\begin{align}
\mu(\boldsymbol{x}) &= \overline{\boldsymbol{u}}  \\
D_p(\boldsymbol{x}) &=\boldsymbol{u}'^2 \tau %\text{d} t
\end{align}
where $\tau$ is a characteristic time of the flow. 
%\begin{equation}
%\mu(\boldsymbol{x}) = \frac{<\text{d} \boldsymbol{x}_k>}{\text{d} t} = <\boldsymbol{u}> = \overline{\boldsymbol{u}}
%\end{equation}
%\begin{equation}
%\sigma^2(\boldsymbol{x}) = \frac{<(\text{d} \boldsymbol{x}_k - <\text{d} \boldsymbol{x}_k>)^2>}{\text{d} t} = <(\boldsymbol{u} - <\boldsymbol{u}>)^2> \text{d} t =  \overline{\boldsymbol{u}'} \text{d} t
%\end{equation}
%BC Eq.~\ref{eq:FP} is numerically integrated in space and time in the computational domain of the non-reacting flow simulation. %  LES using the same numerical schemes for the non-reacting LES.
%Eq.~\ref{eq:FP} is discretized in space directly over the CFD grid (or a coarser version of the mesh used for the non-reacting simulation).
%Equation~\ref{eq:FP} is written and solved for a fixed kernel size $r$. 
%BC The probability density function is normalized so that at each time $t$: 
%BC\begin{equation}
%P_{tot}(t) = \int_V \int_r p(\boldsymbol{x},r,t) \; dr \; dv = 1
%BC\int_V \int_r p(\boldsymbol{x},r,t) \; dr \; d\boldsymbol{x} = 1
%BC\end{equation}
%where the integral over the kernel radius space includes the failure and success probabilities.
In Eq.~\ref{eq:FP} the source term $\dot{Q}_i $ accounts for the transfer between sections due to kernel growth and shrinking.
The kernel growth rate is associated to the local turbulent flame speed $S_T(\boldsymbol{x})$, while the kernel shrinking is driven by the turbulent diffusivity $D_{th,turb}(\boldsymbol{x})$.
%The upper bound of the last section is the critical kernel radius above which the ignition is considered to be a success (or at least the first two phases of the complete ignition process). This value is more difficult to evaluate, but it corresponds to a kernel size that cannot be quenched by the largest eddies of the flow.
%Félix: Related to $l_t$?
The transfers between two neighboring sections during a time interval $\delta_t$ then write:
\begin{equation}
%T_{G,S_{i}\rightarrow S_{i+1}}(\boldsymbol{x},t) = \int_{r^{upp}_{i} - \delta_{G,r}}^{r^{upp}_{i}} p_{i}(\boldsymbol{x},t)dr
T_{G,S_{i}\rightarrow S_{i+1}}(\boldsymbol{x},t) = p_{i}(\boldsymbol{x},t) S_T(\boldsymbol{x})  \delta_t 
\end{equation}
\begin{equation}
%T_{S,S_{i}\rightarrow S_{i-1}}(\boldsymbol{x},t) = \int_{r^{low}_{i}}^{r^{low}_{i} +\delta_{S,r}} p_{i}(\boldsymbol{x},t)dr
T_{S,S_{i}\rightarrow S_{i-1}}(\boldsymbol{x},t) = p_{i}(\boldsymbol{x},t) \frac{D_{th,turb}(\boldsymbol{x})}{\overline{r}_i}  \delta_t 
\end{equation}
where $\overline{r}_i$ is the mean kernel radius in section $S_i$.
%Félix: Il manquait la formule de St. Je reprend ce qui tu avais mis dans ta thèse:
The turbulent flame speed is evaluated following~\cite{Abdel-Gayed:1987,Cordier:2013b}:
\begin{equation}
S _ { T } = S _ { L } ^ { 0 } + n \left( \frac { u ^ { \prime } } { S _ { L } ^ { 0 } } \right) ^ { c } \cdot S _ { L } ^ { 0 }
\end{equation}
where $n$ and $c$ are  model constants from~\cite{Cordier:2013b}. Even if developed in the context of premixed flames, this expression is also used for $NP$ and $SP$ cases as considering an enhancement of the consumption speed by turbulence is still meaningful. Note that $S_L^{tp}$ \cite{Rochette:2018} is used instead of $S_L^0$ in $SP$ case. 

The source term for each section then depends on the time after deposit $t$ and the local flow properties:
\begin{itemize}
\item for $t < t_{CD}$, kernels are only growing and the net change of $p_{i}(\boldsymbol{x},t)$ during a time interval $\delta_t$ is given by:
\begin{equation}
\dot{Q}_i = T_{G,S_{i-1}\rightarrow S_{i}} - T_{G,S_{i}\rightarrow S_{i+1}}
\end{equation}
\item for $t >= t_{CD}$, if $Ka>Ka_c$ kernels are shrinking due to turbulence and the source term writes:
\begin{equation}
\dot{Q}_i = T_{S,S_{i+1}\rightarrow S_{i}} - T_{S,S_{i}\rightarrow S_{i-1}}
\end{equation}
On the contrary if $Ka<Ka_c$, kernels located in flammable mixtures will grow while those located in non-flammable mixtures will shrink:
\begin{align}
%Félix: Peut être dire un mot du pourquoi F_f et  (1-F_f). ?
\dot{Q}_i = & F_f(\boldsymbol{x}) ( T_{G,S_{i-1}\rightarrow S_{i}} - T_{G,S_{i}\rightarrow S_{i+1}} ) \nonumber \\
 & + (1 - F_f(\boldsymbol{x}) ) ( T_{S,S_{i+1}\rightarrow S_{i}} - T_{S,S_{i}\rightarrow S_{i-1}} )
\end{align}
\end{itemize}

Below a minimum size $r_f$ with probability $p_{f}(\boldsymbol{x},t)$, ignition is considered failed. $r_f$ is approximated by the laminar flame thickness $\delta_l^0$ at stoichiometry in the $NP$ and $SP$ cases or at the mixture equivalence ratio for the $P$ case. On the other end above a critical size $r_s$ with probability $p_{s}(\boldsymbol{x},t)$, the flow can no longer extinguish the flame kernel and ignition is successful. This critical size is taken equal to the integral length scale of the turbulent flow, corresponding here to $R_{ext}$, the outer radius of the SWJ at the inlet plane. Note that additional success criteria, such as requiring that the flow direction must be directed toward the injector, could be introduced to generalize the model to other type of configurations, but these were not critical in the present case.
%Kernels in the first section moving to the failed probability or in the last section moving to the success probability, are removed from the calculation and added to $p_{f}(\boldsymbol{x},t)$ and $p_{s}(\boldsymbol{x},t)$, respectively.

The set of $N_{seq}$ Eqs.~\ref{eq:FP} is discretized over an unstructured grid similar to the one used to perform the non-reacting LES but note that because the time-average statistic fields are smoother than the instantaneous LES simulation, a coarser mesh could be used. The equations are integrated using a third-order in space and time two-step Taylor Galerkin scheme \cite{Colin:2000} for the advective term while the diffusive term is solved with a second-order finite element scheme. The equations are advanced in time following an explicit CFL constraint based on  $\mu(\boldsymbol{x})$. A CFL of 0.7 used in all the results presented hereafter. A set of Eqs.~\ref{eq:FP} is numerically integrated for each sparking location $\boldsymbol{x_0}$. Starting from the initial kernel, all $p_{i}(\boldsymbol{x},t)$ -except the one corresponding to the initial kernel size- first increase progressively, before decreasing down to zero at the end of the simulation, when all kernels have reached either a quenched or ignited state. Therefore only $p_{f}(\boldsymbol{x},t)$ and $p_{s}(\boldsymbol{x},t)$ end with non-zero values, $p_f^{end}(\boldsymbol{x},\boldsymbol{x_0})$ and $p_s^{end}(\boldsymbol{x},\boldsymbol{x_0})$ respectively, and the probability of successful ignition for sparking at $x_0$ is  simply:
\begin{equation}
P_{ign}(\boldsymbol{x_0})  = \int_V p_{s}^{end}(\boldsymbol{x},\boldsymbol{x_0}) \; dV
\end{equation}

\section{Results}
\label{sec_results}

The model is now applied to the three operating conditions listed in Table \ref{Tbl:expe_cond}. The model parameters used for each case are listed in Table \ref{tab:model_param}. The choice of the number of sections was motivated by the observation that in most cases studied here, the kernel radius distribution featured a single peak, which can be well reproduced with a relatively low number of sections. Note that the computational cost of the model is directly proportional to the number of sections.

\begin{table}
\caption{Summary of the model physical and numerical parameters}
\centering
\begin{tabular}{| l | c c c |}
\hline
  & $P$ & $NP$  & $SP$  \\
\hline
\hline 
r$_{s}$ [m] & 0.01 & 0.01 & 0.008  \\ 
\hline 
r$_{f}$ [m] & 0.001 & 0.0008 & 0.001  \\
\hline
$Ka_c$ & 4.5 & 4.5 & 4.5 \\ 
\hline
$N_{seq}$ & 12 & 12 & 12 \\
\hline
\end{tabular}
\label{tab:model_param}
\end{table}

\subsection{Ignition probability maps} \label{sec:ignit_prob_maps}
%Comparison of the model results versus experimental ignition maps for all three cases
% This section should harmonized and presented in a more progressive manner: cases of increasing complexity, a different 'physics' affects the kernel development for each cases.

%% Premixed and non-premixed cases

\begin{figure}[ht!]
\centering
\includegraphics[width=0.48\textwidth]{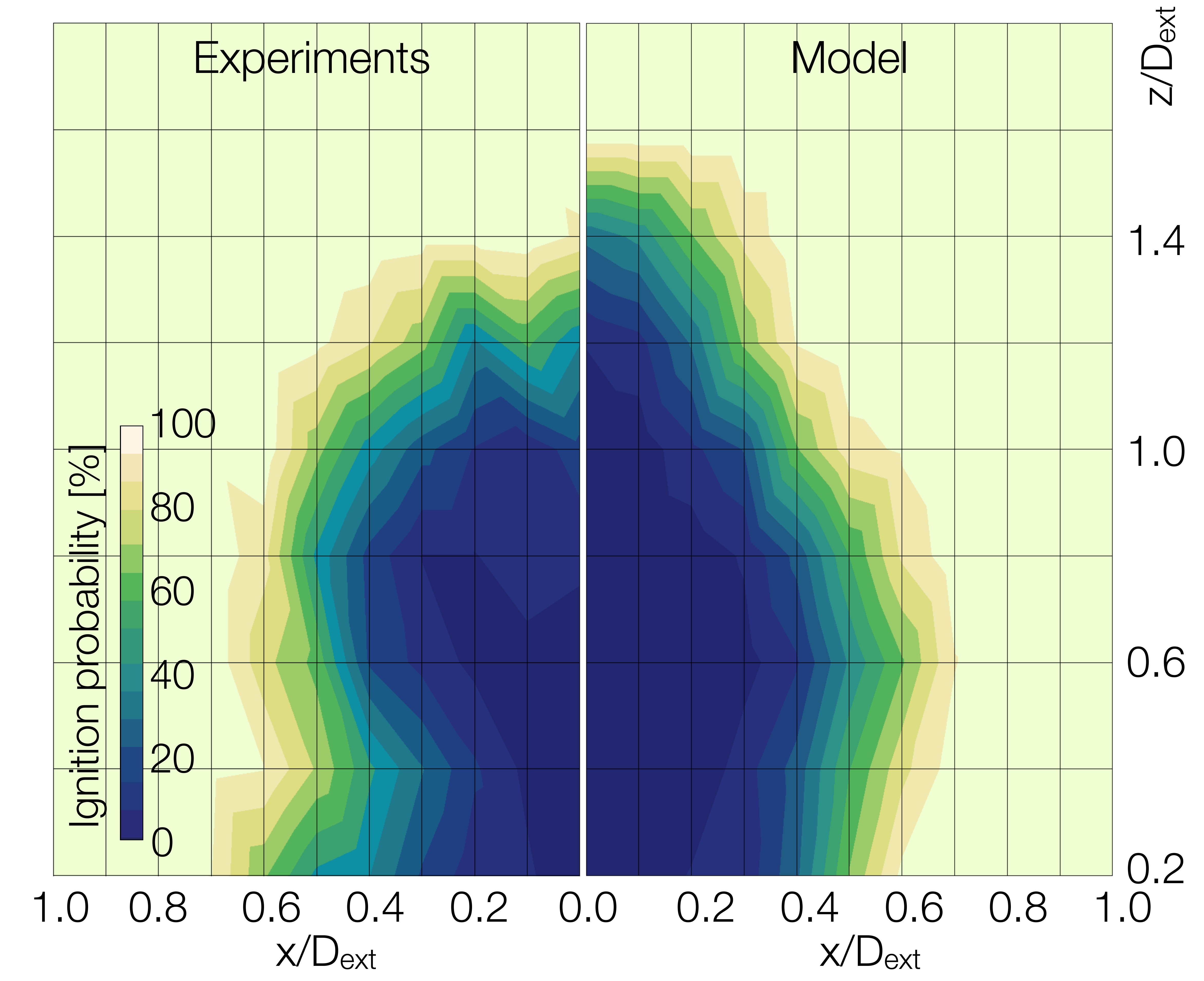}
\caption{$P$ case. Comparison between experimental (left) and MIST (right) ignition probability maps in the solid line box of Fig.~\ref{fig:stream}.}
\label{fig:Pmap}
\end{figure}

\begin{figure}[ht!]
\centering
\includegraphics[width=0.48\textwidth]{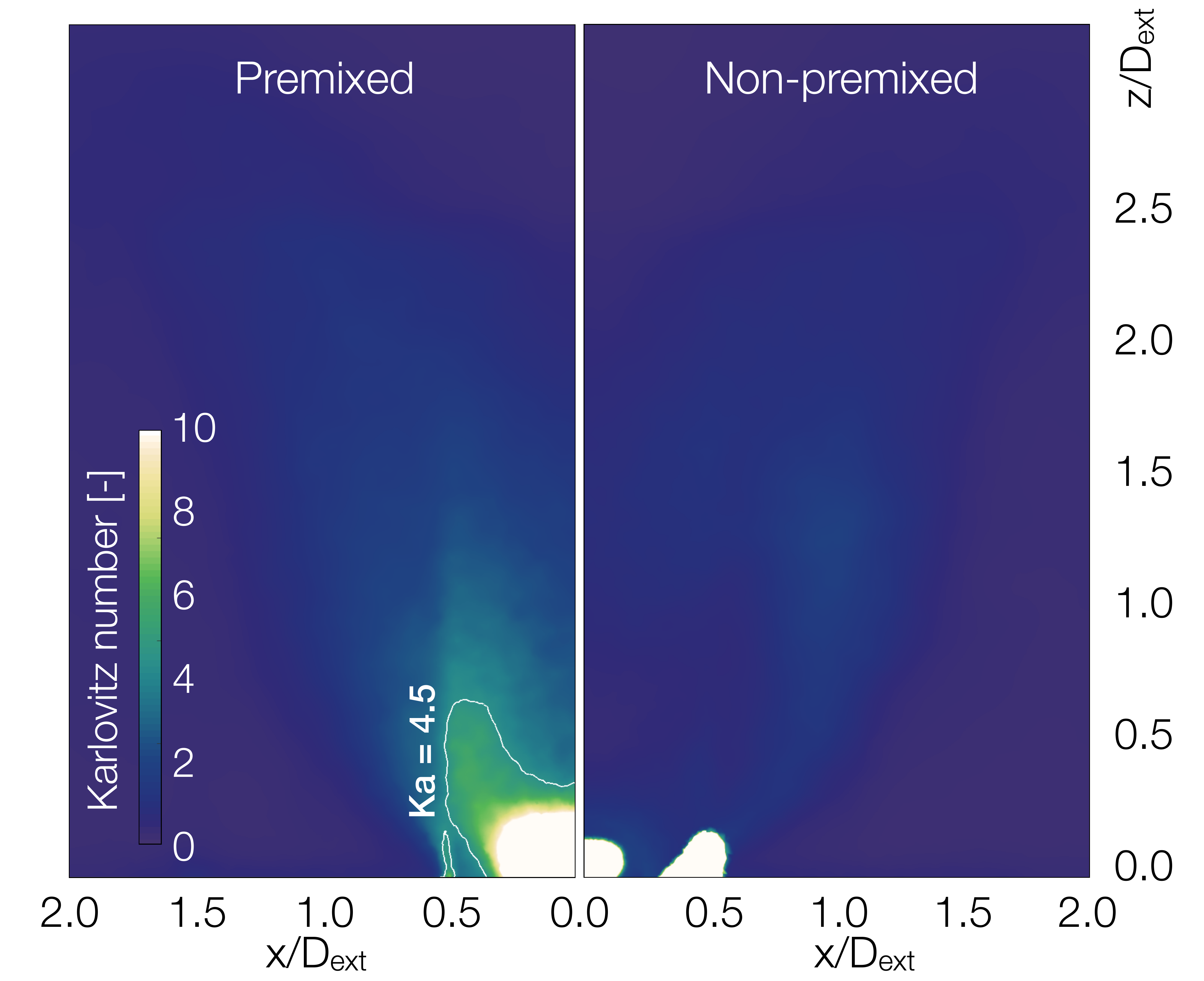}
\caption{Karlovitz number (Eq.~\ref{eq:Ka_abdel}) contours in a central cut plane through the computational domain in the $P$ (left) and $NP$ (right) cases.}
\label{fig:Karlovitz}
\end{figure}

The results obtained with MIST for case $P$ are compared to the experiment in Fig.~\ref{fig:Pmap}. The map corresponds to the solid line box in Fig.~\ref{fig:stream}. The shape of the ignition probability distribution predicted by MIST is in fairly good agreement with the experiment. A large region of low ignition probability is found along the central axis up to an axial position of $z/D_{ext} = 1.4$, which globally follows the limits of the IRZ. In this premixed case flame stretch is the only quenching mechanism, illustrated in Fig.~\ref{fig:Karlovitz}(left): the Karlovitz number exceeds the critical value $Ka_c = 4.5$ only in the IRZ close to the injection. The low ignition probability is therefore the result of recirculating kernels in the IRZ, subjected to varying but high flame stretch for a long time. Aside from this central region, the ignition probability is 1 everywhere.

The differences between MIST and the experiments are mostly concentrated in the transition between the low and high ignition probability regions, with sharper gradient observed in the model results. This can be expected from the model formulation which predominantly follows the mean kernel trajectory whereas intermediate ignition probability often results of equally probable kernel paths (two or more) which can differ significantly from the mean. Additionally, this case was found to be the most sensitive to the choice of $Ka_c$: value of $Ka_c$ below 2.0 resulted in an over-extended high $Ka$ region encompassing most of the SWJ and the upstream part of the IRZ, and resulting in a wide over-prediction of the low $P_{ign}$ region. With a $4 < Ka_c < 8$, the region of high $Ka$ remains confined close to the stagnation point and results consistent with those of Fig.~\ref{fig:Pmap} were obtained, with the position of the low to high probability transition along the central axis moving downward with increasing $Ka_c$.

\begin{figure}[ht!]
\centering
\includegraphics[width=0.48\textwidth]{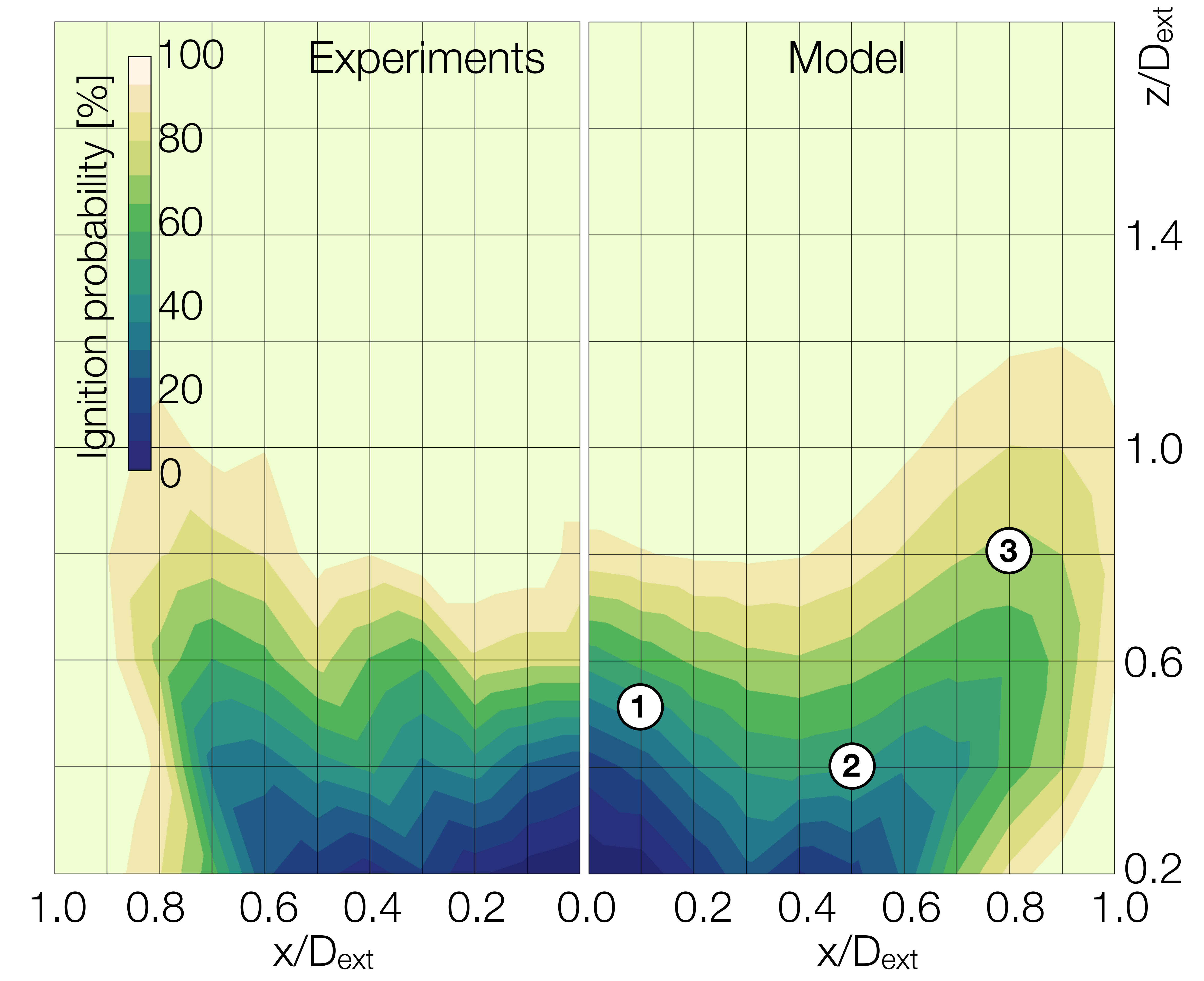}
\caption{$NP$ case. Comparison between experimental (left) and MIST (right) ignition probability maps in the solid line box of Fig.~\ref{fig:stream}.}
\label{fig:NPmap}
\end{figure}

The $NP$ case results are now compared to experiment in Fig.~\ref{fig:NPmap}. Again a good agreement is observed, and in both maps low ignition probability regions are found close to the methane central jet and in the wake of the air SWJ. Contrary to case $P$, the region of high Karlovitz number is very small (Fig.~\ref{fig:Karlovitz}) due to the near stoichiometric conditions in the lower part of the IRZ. In fact the shape of low ignition probability regions closely follow the flammability factor distribution depicted in Fig.~\ref{fig:Mixing}: ignition is mainly controlled by mixing. %However, the analysis of the model results and the LES of ignition \cite{Esclapez:2015} shows that ignition can occur even with locally detrimental conditions at the spark location.

%%
%% Spray case

\begin{figure*}[h!]
\centering
\includegraphics[width=0.6\textwidth]{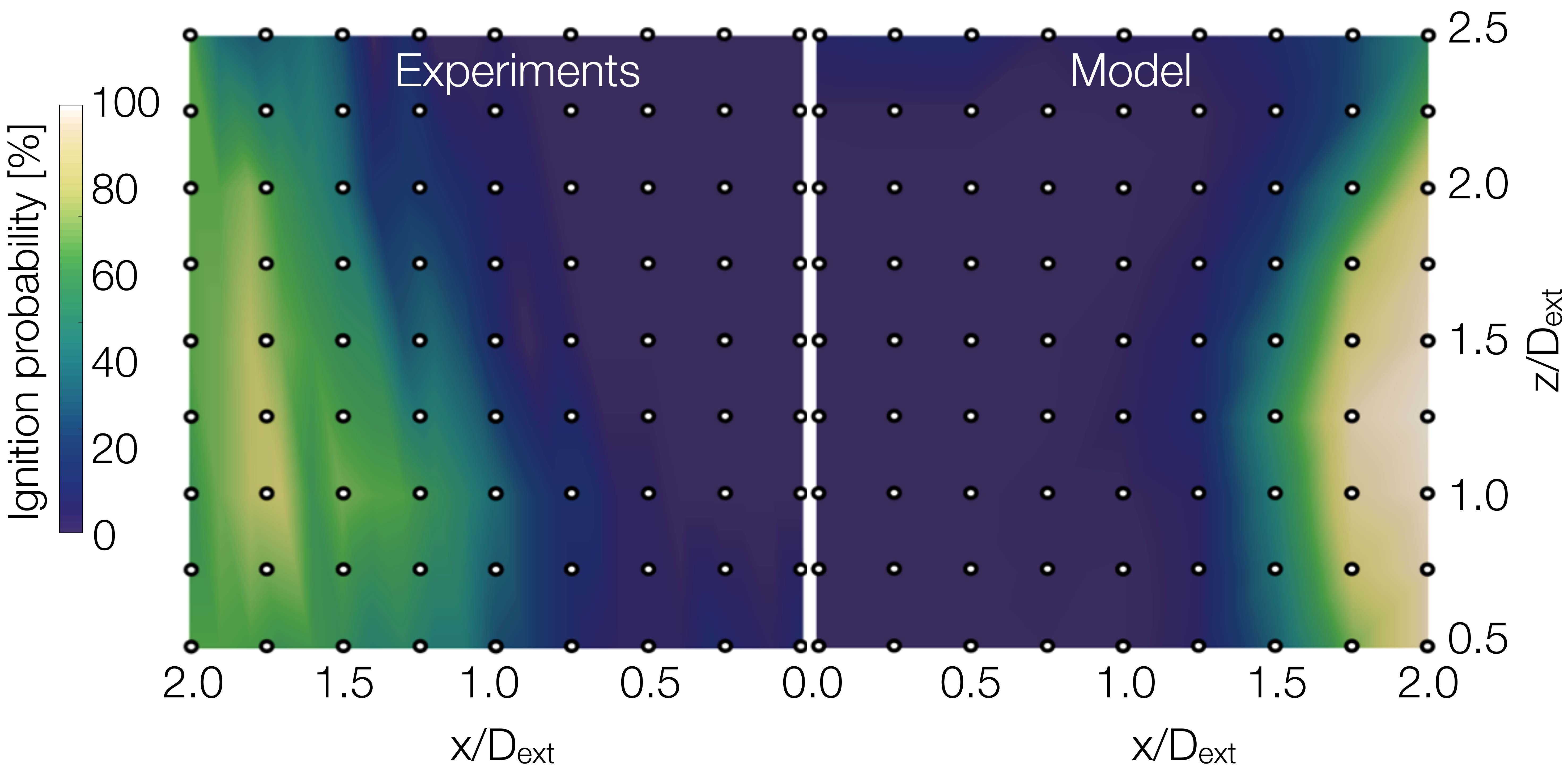}
\caption{$SP$ case. Comparison between experimental (left) and MIST (right) ignition probability maps in the dashed box of Fig.~\ref{fig:stream}.}
\label{fig:carto_spray}
\end{figure*}

Finally the comparison with experiment is made for case $SP$ in Fig.~\ref{fig:carto_spray}, in the dashed box of Fig.~\ref{fig:stream}. The agreement is again quite satisfactory. The same overall topology of the ignition probability map is recovered. The entire IRZ is characterized by very low ignition probability, below $0.1$, and the CRZ is the most ignitable region of the chamber, with ignition probability above $0.7$ near the lateral wall. Finally the gradient of $P_{ign}$ more or less coincides with the SWJ, slightly shifted in MIST by around $0.5~D_{ext}$ towards the CRZ. %Still, the gradient is correctly inclined, following the SWJ. 

\begin{figure}[h!]
\centering
\includegraphics[width=0.48\textwidth]{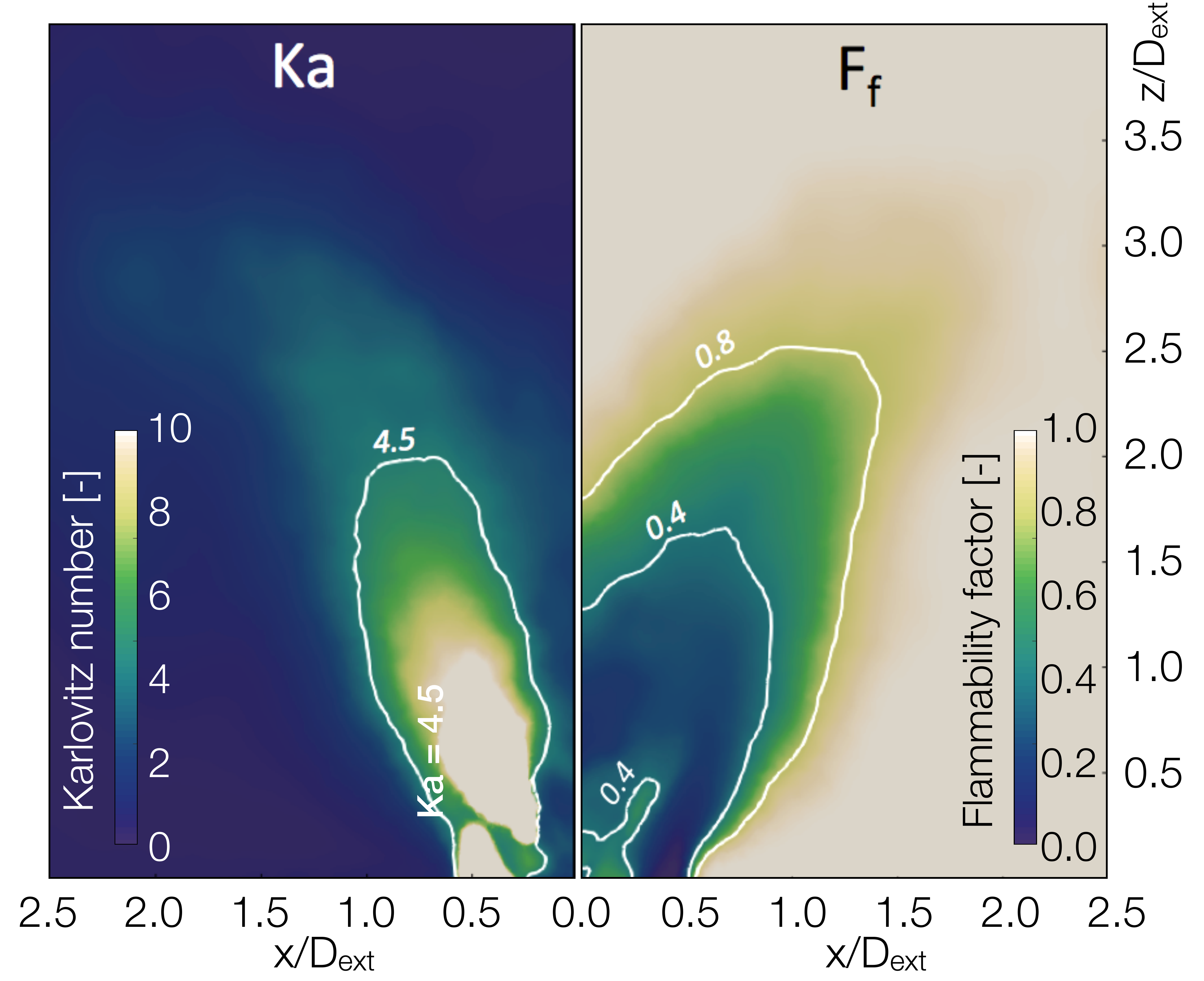}
\caption{$SP$ case. Maps of Karlovitz number (left) and flammability factor (right).}
\label{fig:Ka_Ff_spray}
\end{figure}

This topology of $P_{ign}$ is strongly related to local non-reacting flow properties $Ka$ and $F_f$ shown in Fig.~\ref{fig:Ka_Ff_spray}. The very homogeneous flammable mixture combined with a low Karlovitz number (due to low velocity fluctuation levels) found in the CRZ explain the very high ignition probability. On the contrary, the IRZ and the bottom of the SWJ are very lean with high velocity fluctuations, leading to high local Karlovitz number above the critical value $Ka_c > 4.5$. 

\subsection{Detailed analysis}
\label{ssec:LES_R}

\subsubsection{Premixed case}

\begin{figure*}[h!]
\centering
\includegraphics[width=1.0\textwidth]{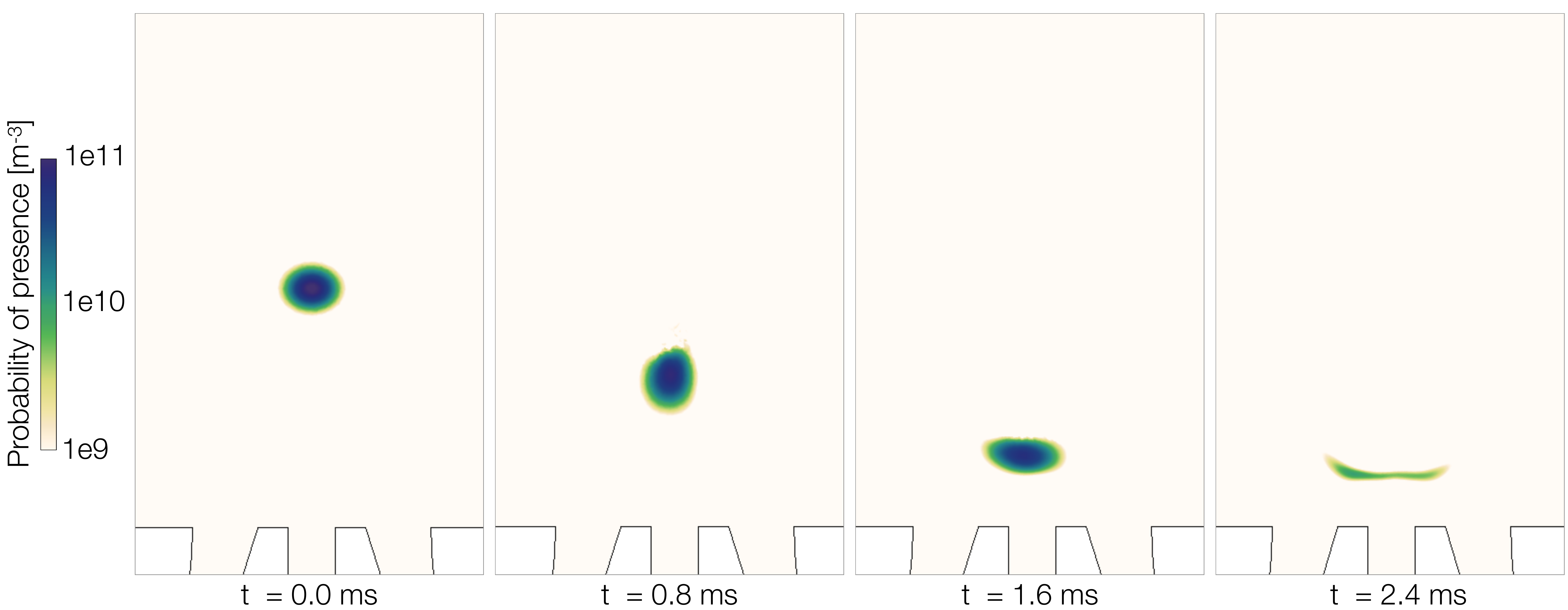}
\caption{$P$ case. Probability density of presence $p(\boldsymbol{x},t)$ of all size kernels in a central cut-plane at four instants for sparking at $(r/D_{ext} = 0.0,z/D_{ext} = 1.0)$.}
\label{fig:prem_transient1}
\end{figure*}

To illustrate the capabilities of MIST to correctly reproduce the time evolution of kernels,  the temporal evolution of kernels of all sizes $P_{pres}(\boldsymbol{x},t)=\int  P_{pres}(\boldsymbol{x},r,t)dr$ is shown in Fig.~\ref{fig:prem_transient1} for a sparking location at $(r/D_{ext} = 0.0, z/D_{ext} = 1.0)$ where both experiment and MIST indicate that the ignition probability is close to 0$\%$ (Fig.~\ref{fig:Pmap}). At this location, the mixture is flammable and the low level of turbulence results in $Ka < Ka_c$. However the recirculating mean flow rapidly entrains most kernels towards the high Karlovitz region near the injection system, before they reach a sufficient size to resist the strong local turbulence there. This reflects in the motion of the peak $P_{pres}$ towards the injection system, where it finally vanishes. This behavior is consistent with the ignition failure mechanism observed both experimentally and numerically \cite{Cordier:2013, Barre:2014}.

\begin{figure*}[h!]
\centering
\includegraphics[width=1.0\textwidth]{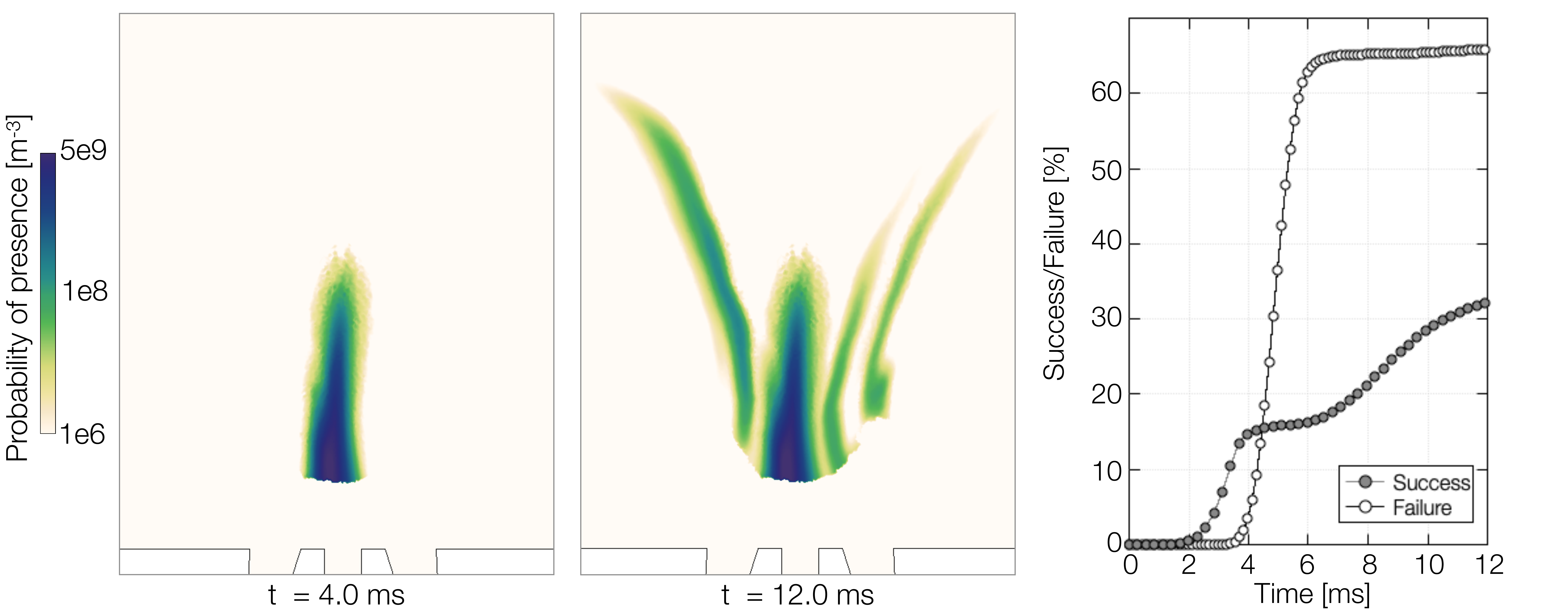}
\caption{$P$ Case. Probability density of successful ignition, $p_{s}(\boldsymbol{x},t)$ in a central cut-plane at two instants for sparking at $(r/D_{ext} = 0.0,z/D_{ext} = 1.4)$, and temporal evolution of $P_s(t)=\int_V p_{s}(\boldsymbol{x},t) dV$ and $P_f(t)=\int_V p_{f}(\boldsymbol{x},t) dV$.}
\label{fig:prem_transient2}
\end{figure*}

Intermediate values of the ignition probability found at the limit of the IRZ, correspond to an increased proportion of kernels that have time to reach a sufficient size before entering the high $Ka$ region. Two scenarios are observed: 1) a fast ignition scenario where the kernel grows fast and leads to ignition while in the IRZ,  2) a delayed ignition scenario where the kernel growth is sufficient to avoid extinction in the high $Ka$ region, but not to  ensure ignition there, which then occurs later in the SWJ. The existence of these two ignition modes is clearly visible in Fig.~\ref{fig:prem_transient2} illustrating ignition in the central cut-plane of the burner when sparking at $(r/D_{ext} = 0.0,z/D_{ext} = 1.4)$ where experimental $P_{ign}$ is 32\%. At $t = 4$ ms, $P_{ign} \simeq 15$ \% and the zone of high ignition success probability density $p_{s}(\boldsymbol{x},t)$ corresponds to upstream kernel trajectories inside the IRZ, i.e., the first scenario. Later at $t = 12$ ms the zone extends along trajectories in the SWJ, indicating delayed ignition of the second scenario. The temporal evolutions of $P_s(t)=\int_V p_{s}(\boldsymbol{x},t) dV$ and $P_f(t)=\int_V p_{f}(\boldsymbol{x},t) dV$ show as well the two modes, with a first increase of $P_s$ around $3$ ms, followed by a plateau before a second increase starting later around $7$ ms.   
These results highlight the ability of MIST to capture non-monotonic evolutions of the kernel size as its trajectory successively enters regions that promote or impede its growth.

\subsubsection{Non-premixed case}

The ignition probability at three locations (shown in Fig.~\ref{fig:NPmap}) was directly computed by performing 20 LES of ignition in a previous study \cite{Esclapez:2015}. Table~\ref{tab:LES_XP_datas} reports the ignition probability obtained from experiment, LES and MIST. Both LES and MIST give very similar results, also close to measurements. Note that about 5 million CPU hours have been required for each data point with LES whereas it took only few minutes with MIST.

\begin{table}[ht!]
\centering
\begin{tabular}{l c c c}
\hline
 	& Exp. 		& LES \cite{Esclapez:2015}	& MIST 	\\\hline
 %	& $P_{ign,XP}$	& $P_{ign,LES}$   			& $P_{ign,MIST}$  \\\hline
PT1	& 28-70\% 	& 40\%                           		& 38\%             \\
PT2	& 50\%      	& 48\%                        		& 50\%                \\
PT3	& 80\%      	& 72\%                       		& 74\%                \\
\hline
\end{tabular}
\caption{$NP$ case. Comparison of $P_{ign}$ from experiment \cite{Cordier:2013}, LES and MIST at the three sparking locations 1, 2 and 3 shown in Fig.~\ref{fig:NPmap}.}
\label{tab:LES_XP_datas}
\end{table}

\begin{figure*}[h!]
\centering
\includegraphics[width=1.0\textwidth]{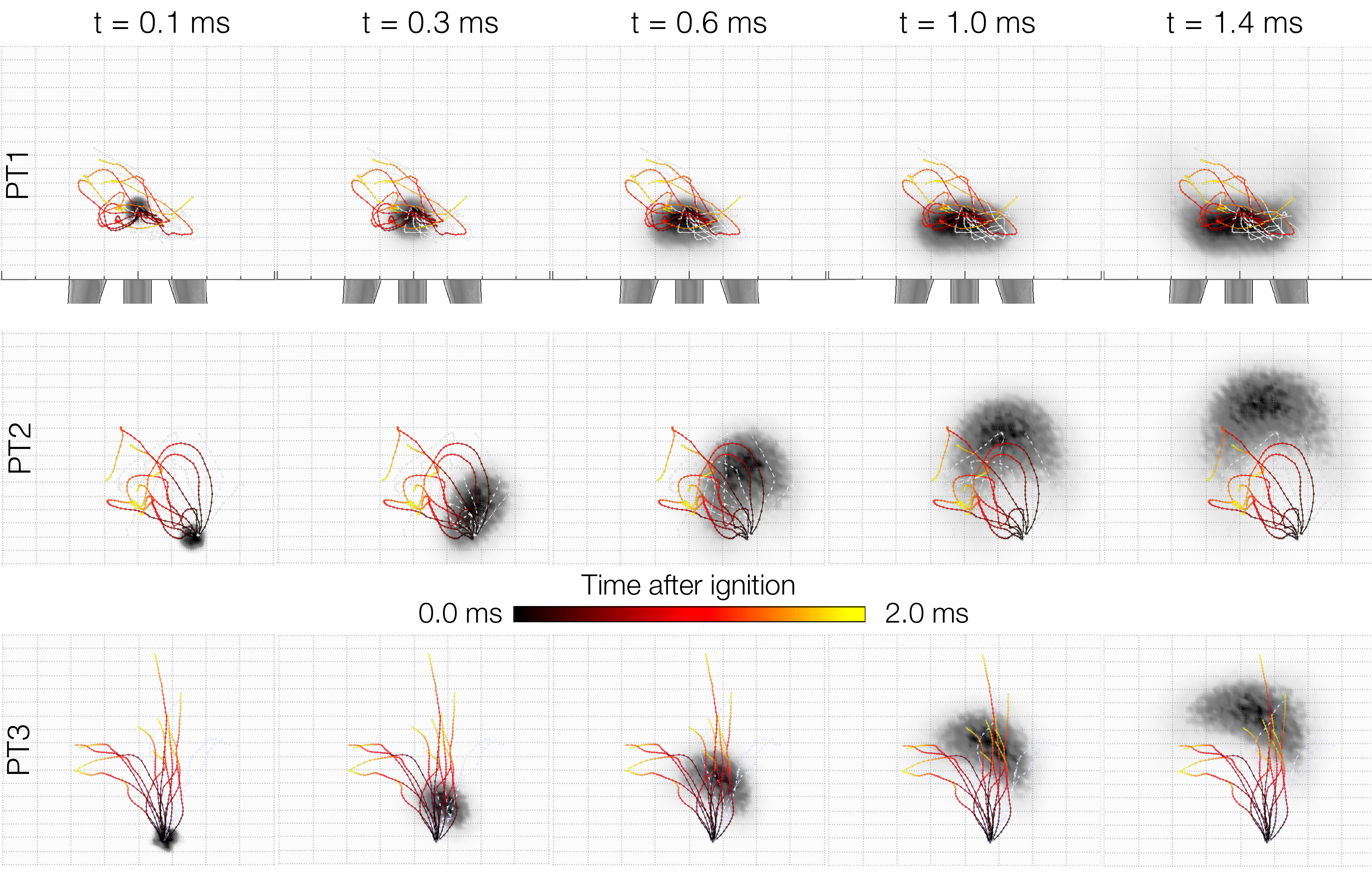}
\caption{$NP$ case. Two-dimensional projection of $p(\boldsymbol{x},t)$ of all size kernels obtained from MIST (grayscale) with overlaid kernel trajectories obtained from LES (lines) colored by the time after ignition.}
\label{fig:compaLESMIST}
\end{figure*}

To analyze deeper the ignition scenarios, kernel trajectories are extracted from LES where each kernel is represented by the center of gravity of the volume defined by $T > 1300$ K.
%\begin{equation}
%\boldsymbol{x_{k}} = \sum_{n=1}^{N} \frac{\rho_n V_n \boldsymbol{x}_n}{\rho_n V_n}
%\end{equation}
%where $rho_n$, $V_n$ and $\boldsymbol{x}_n$ are the density, volume and coordinates of vertex $n$, and $N$ is the total number of vertices above 1300 K. 
Both LES trajectories and the MIST PDF of presence  $p(\boldsymbol{x},t)$ of all size kernels are projected on 2D-maps for the three sparking locations in Fig.~\ref{fig:compaLESMIST}. LES trajectories are colored with time to compare with the time evolution of $p(\boldsymbol{x},t)$.

Results indicate that MIST qualitatively agrees with LES and is able to reproduce the different kernel motion trends associated with each sparking location: 
\begin{itemize}
\item at PT1, the flame kernel first stays close to the stagnation point (until $\approx 1$ ms) and is eventually convected along the SWJ for successful events,
\item at PT2, the sparking in the shear layer between the IRZ and the SWJ leads to two categories of kernel trajectories, either along the SWJ or trapped in the IRZ,
\item at PT3, all trajectories mainly follow the SWJ, going downstream and rotating around the nozzle axis.
\end{itemize}
However it also highlights some limitations of the model. At the vicinity of PT1, although both LES and experiments have shown significant deformation and fragmentation of the kernel, MIST assumes that the kernel remains spherical. This difference can partially explain the wide range of instantaneous kernel trajectories observed in the LES, which is not captured by the dispersion of the trajectories in MIST. 

\subsubsection{Spray case}

\begin{figure*}[h!]
\centering
\includegraphics[width=0.95\textwidth]{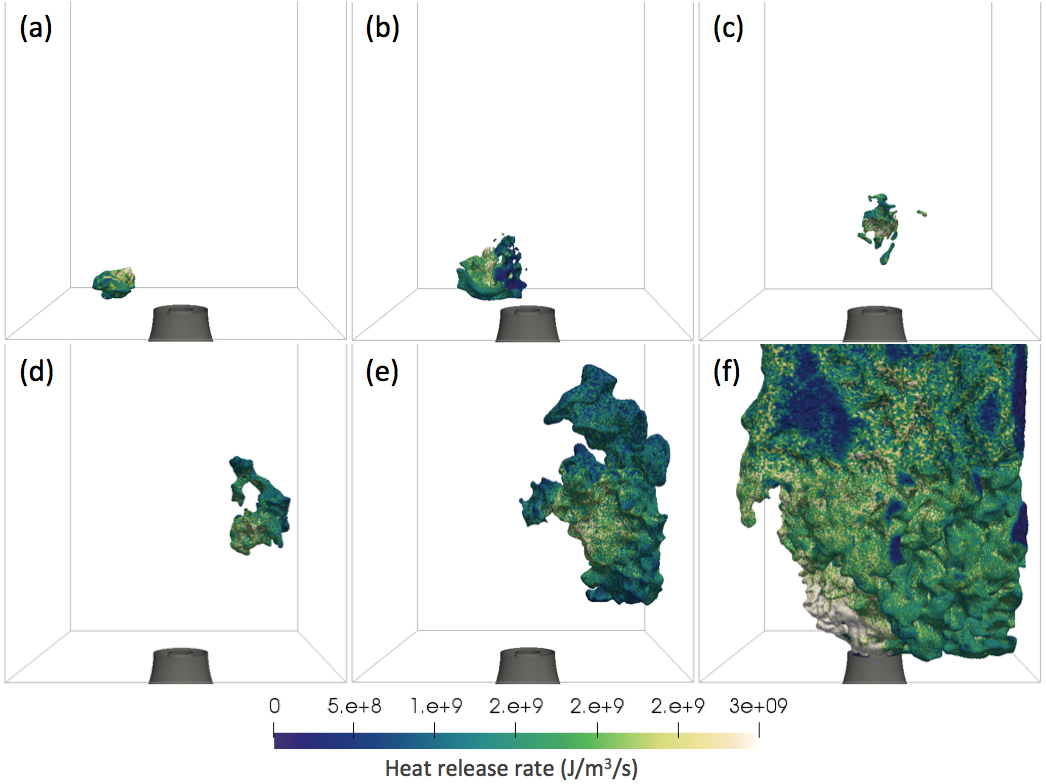}
\caption{$SP$ case. Snapshots of LES for sparking at $(r/D_{ext} = 1.5,z/D_{ext} = 0.5)$. Flame front visualization (iso-$T=1500$ K) colored by heat release rate. Time after spark: (a) $1.7$ ms, (b) $3.7$ ms, (c) $7.7$ ms, (d) $13$ ms, (e) $16$ ms, (f) $20$ ms.}
\label{fig:LES_spray}
\end{figure*}

%As explained in Sec.~\ref{sec:ignit_prob_maps}, $P_{ign}$ and $P_{ker}$ are sometimes very different and the final outcome of an ignition sequence is not only determined by properties at the spark location. To illustrate how MIST recovers such complex ignition behaviour in the $SP$ case, 
As for case $NP$ in the previous section, MIST is compared to LES of ignition sequences, at the sparking location $(r/D_{ext} = 1.5, z/D_{ext} = 0.5)$. The experimental ignition probability found at this position is $50~\%$. Snapshots of the flame front (iso-$T=1500$ K) colored by the heat release rate are given in Fig.~\ref{fig:LES_spray} at different times after the spark, extracted from the LES of a successful ignition. Starting from the bottom of the CRZ, the kernel is first convected towards the injector by the recirculating flow (a). During this phase, the kernel grows as it meets favorable conditions. When arriving above the air inlet (b), the flame kernel subjected to very high velocity fluctuations, may rapidly quench. The kernel is then convected downstream by the SWJ (c) and is still strongly shredded in this turbulent zone. If able to survive, the kernel finally reaches the much favorable top part of the CRZ (d) after $10$ ms, where it grows fast to extend over the entire CRZ and the SWJ (e), and eventually ignites the full chamber. In this late ignition scenario the kernel convection plays a critical role. %Many other ignition scenarios have been observed from the same sparking position: fast ignition in the CRZ when the residence time in this zone increases, fast misfire in the bottom part of the SWJ if this residence time is too small, or late misfire in the top part of the SWJ.  

\begin{figure*}[h!]
\centering
\includegraphics[width=0.3\textwidth]{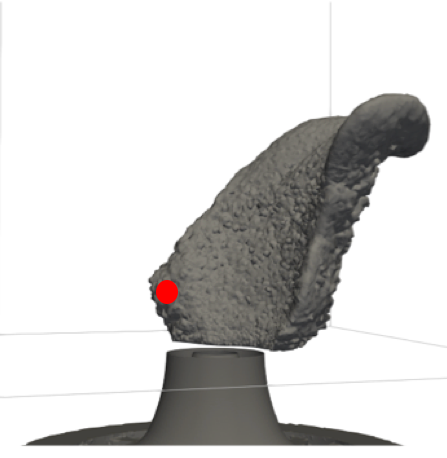}
\hspace*{2cm}
\includegraphics[width=0.3\textwidth]{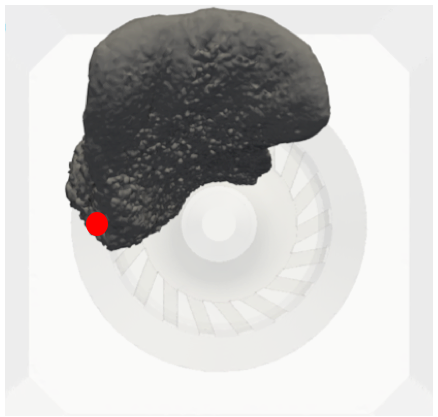}
\caption{$SP$ case. Prediction of MIST for sparking at $(r/D_{ext} = 1.5,z/D_{ext} = 0.5)$ (red dot) : final ($t ~$ 12 ms) iso-surface of all positions where $r_s$ was reached. Left: side view; Right: top view.}
\label{fig:spray_isoC_rk}
\end{figure*}

\begin{figure*}[h!]
\centering
\includegraphics[width=0.95\textwidth]{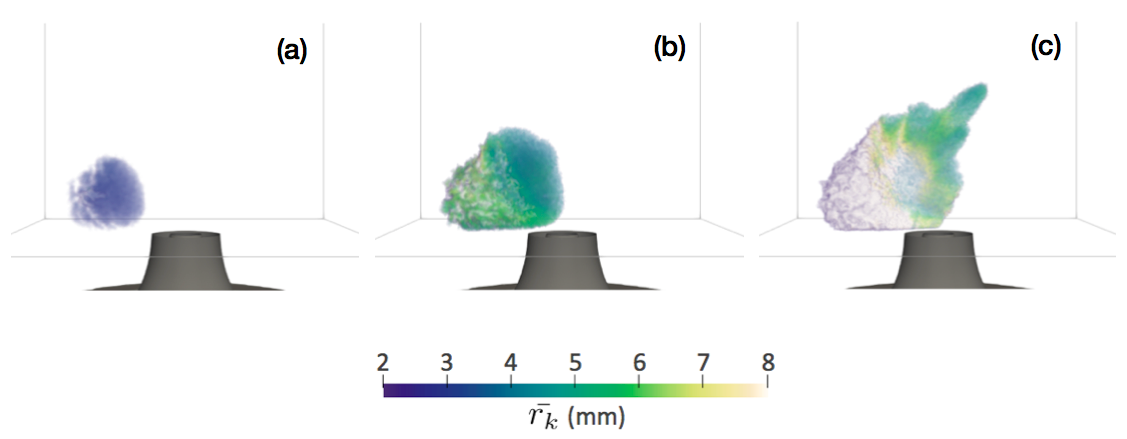}
\caption{$SP$ case. Volume rendering of the mean flame kernel radius after $1$ ms (a), $3$ ms (b) and $7$ ms (c). }
\label{fig:spray_rkmean}
\end{figure*}

The above LES sequence is to be compared with the prediction of MIST, illustrated in Figs.~\ref{fig:spray_isoC_rk} and~\ref{fig:spray_rkmean}. MIST predicts at this point an ignition probability of $40~\%$, close to the experimental value of $50~\%$. In Fig.~\ref{fig:spray_isoC_rk}, the cumulated iso-surface of all positions of the chamber where $r_s$ has been reached, independently of the time after spark, is very similar to Fig.~\ref{fig:LES_spray}~(e) showing that MIST is able to reconstruct the ignition scenario.   

Figure~\ref{fig:spray_rkmean} provides a front view of the iso-volume of mean flame kernel radius above $0.01$ mm at three times after sparking. After $1$ ms (a), the kernel convection phase in the bottom part of the CRZ is found similar to the LES ignition sequence (Fig.~\ref{fig:LES_spray}a). At this early time, the mean kernel size is $\approx 3$ mm and progressively increases. After $3$ ms (b), the larger iso-volume indicates a dispersion of the kernel trajectories. Kernels staying longer in the favorable CRZ grow much more than those entering the adverse SWJ. This is demonstrated by $\bar{r_k}$ reaching $7$ mm in the CRZ while remaining below $\approx 5$ mm in the SWJ. The most advanced points of the iso-volume (towards the SJW) correspond to kernels leaving the CRZ most rapidly, thus having the lowest radius near $3$ mm. This is comparable to what can be observed from the LES in  Fig.~\ref{fig:LES_spray}b. Finally after $7$ ms (c), kernels that stay longer in the CRZ reach $r_s = 8$ mm. On the contrary, kernels convected downstream in the SWJ grow more slowly as in Fig~\ref{fig:LES_spray}c and d.  
For this case again, it is remarkable to observe that MIST is able to recover the wide range of flame kernel trajectories and size evolutions, and the balance between kernel growth in favorable regions and kernel destruction by strong turbulence.

\section{Conclusions}
\label{sec_conclusion}

In this work, a model for ignition statistics (MIST) is proposed in order to predict the ignition probability from a non-reacting flow solution. More specifically, MIST aims at predicting the success of creating a sufficiently large, self-sustained flame kernel during the first few milliseconds after energy deposit. MIST differs from previous ignition models in that it directly combines local flame extinction indicators with statistics of the flame kernel trajectories in order to include transient effects due to the flame kernel motion before ignition. In addition MIST does not need to compute multiple independent ignition events to build kernel trajectory statistics, thanks to a fully statistical approach. This allows to drastically reduce the computational cost, down to few minutes to build a full ignition map. The model is tested on an academic swirled burner operated in premixed, non-premixed and two-phase conditions. In all cases, the model is able to reproduce with good accuracy the ignition probability map obtained experimentally. Detailed analysis of the model behavior indicates that MIST provides valuable insights on the ignition success and failure mechanisms, consistent with the behaviors observed from multiple ignition sequences both experimentally and numerically. This good prediction and efficiency performances make MIST a very attractive tool for the optimization of the igniter position and conditions of real aeronautical combustion chambers. Further improvements of the model include a better description of the interactions between the flame kernel and the walls, a critical aspect of spark plug location in practical systems. \\

\noindent {\bf Acknowledgements}

The authors thank M. Cordier, J. Marrero-Santiago, B. Renou and co-workers from CORIA for fruitful collaboration. This work was performed using HPC resources from GENCI-IDRIS (Grant 2013- x20132b5031) and TGCC (allocations  2016153551 and \& A0032B10157 made  by PRACE and GENCI respectively).\\

%\clearpage

\input{journaux-dec10.tex}
%\bibliographystyle{elsarticle-num}
%\bibliography{bib_ignit_model}

\input{Esclapez_CF_MIST.bbl}
\appendix
\section{Validation of the non-reacting flow statistics}
\label{app:valid_cold}

\begin{figure*}[ht!]
\centering
\includegraphics[width=1.0\textwidth]{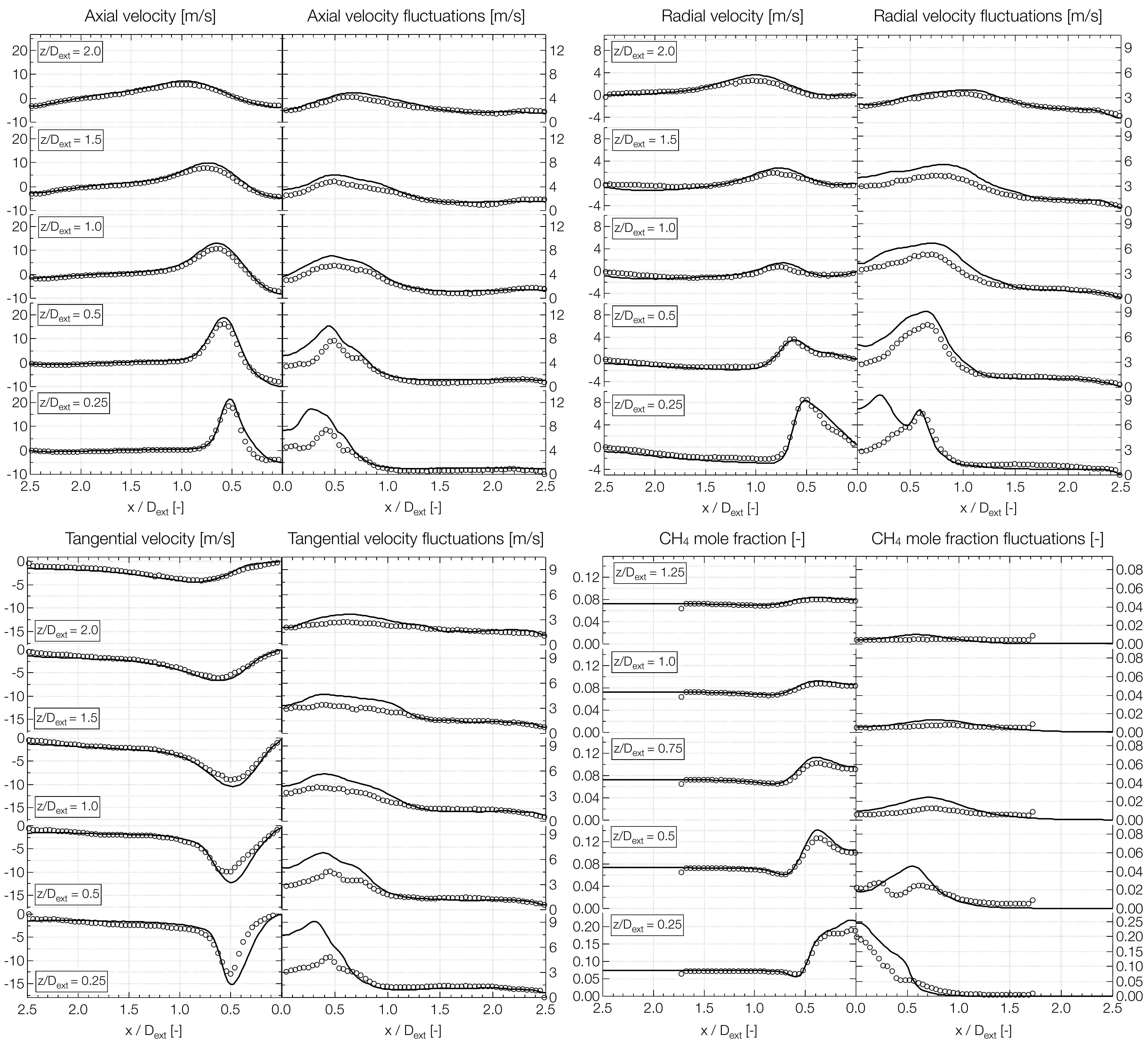}
\caption{Comparison of time-averaged velocity and fuel mole fraction mean and fluctuating profiles from non-reacting LES against experimental data at 5 axial locations.}
\label{fig:NRvalid}
\end{figure*}

Figure~\ref{fig:NRvalid} shows a comparison for the $NP$ case between LES and experimental data at five axial positions downstream of the injector for the three components of velocity and the fuel mole fraction as well as their fluctuations. A very good agreement is observed for all mean quantities while fluctuations are over-predicted close to the injector. Note that this region corresponds to the very fine mesh region where the characteristic grid size is smaller than the PIV window. %Fuel distribution is well reproduce in the LES with a small over-prediction of the fuel mass fraction near the upstream end of the IRZ, where the fuel mole fraction fluctuations are also higher than in the experiments.

\begin{figure*}[h!]

\begin{minipage}{0.5\linewidth}
\centering
\subfloat[]{\label{fig:valid_ptcl_axial} \includegraphics[width=.98\textwidth]{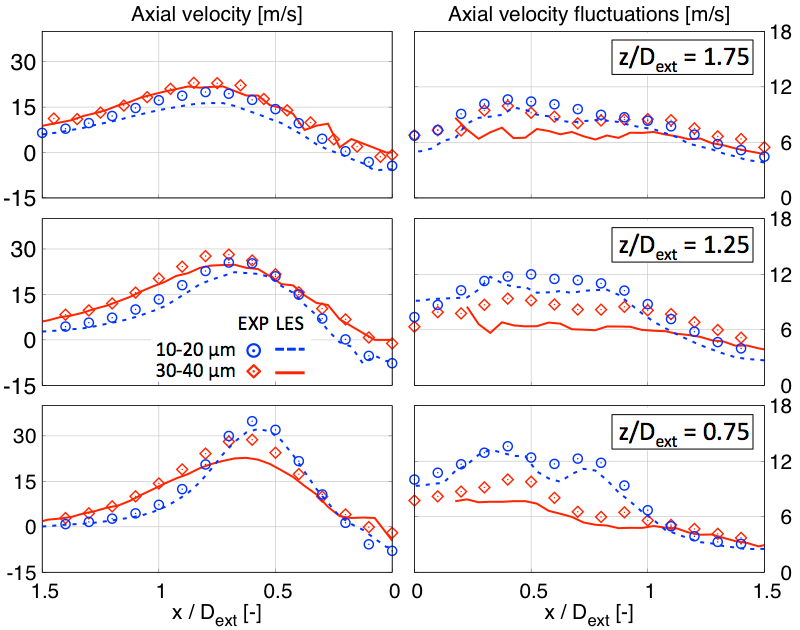}}
\end{minipage}%
\begin{minipage}{0.5\linewidth}
\centering
\subfloat[]{\label{fig:valid_ptcl_radial} \includegraphics[width=.98\textwidth]{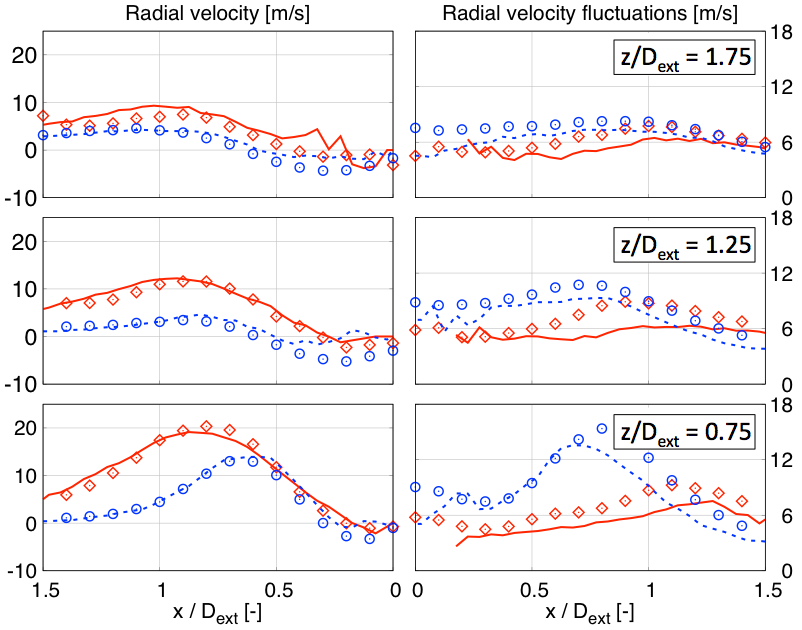}}
\end{minipage}%
\\
\begin{minipage}{1\linewidth}
\centering
\subfloat[]{\label{fig:valid_ptcl_tang} \includegraphics[width=.49\textwidth]{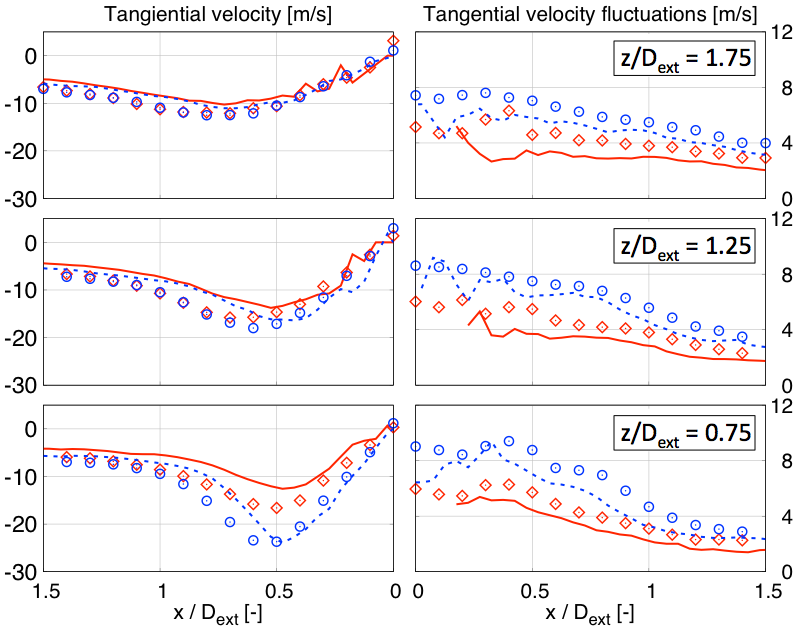}}
\end{minipage}%

\caption{\label{fig:valid_ptcl} Comparison of time-averaged droplet velocity mean and fluctuating profiles from non-reacting LES against experimental data at 3 axial locations.}

\end{figure*}

Figure~\ref{fig:valid_ptcl} shows a comparison for the $SP$ case between LES and experimental data at three axial positions downstream for the droplet velocity. The three components of velocity are shown for two diameter classes: $10-20$ $\mu$m and $30-40$ $\mu$m. A very good agreement is found for both mean and fluctuating values, although the latter are slightly under-predicted. Small droplets are found to align with the carrier phase contrary to larger ones that are much more inertial. This difference between small and large droplet axial velocity however reduces at higher axial locations.

\section{Validation of the mixture composition statistics}
\label{app:mixt_stat}

\begin{figure}[ht!]
\centering
\includegraphics[width=0.48\textwidth]{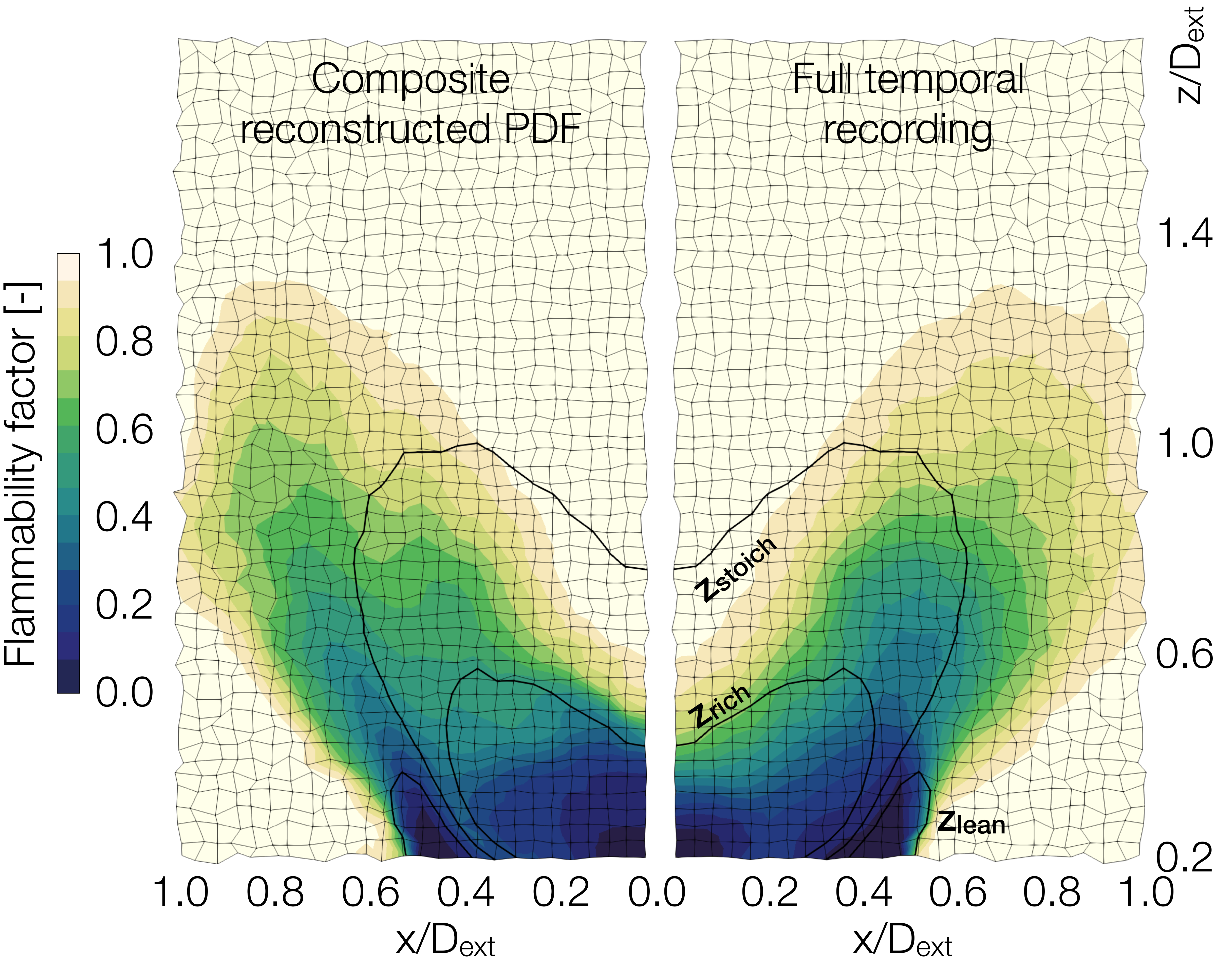}
\caption{Map of flammability factor $F_f$ obtained from MIST (Eq.~\ref{eq:Ffmodel}, left) and LES (right), along with iso-lines of time-averaged lean, rich and stoichiometric mixture fraction.}
\label{fig:flamma}
\end{figure}

In both $NP$ and $SP$ cases, the flammability factor and the mean flammable mixture are reconstructed from time-averaged data. The method proposed in Section~\ref{sssec:mixture} is validated against data extracted from non-reacting LES in the $NP$ case. The simulation is run for $150$ ms during which temporal signals of mixture composition are recorded at 1331 locations in order to map the measurement window (see full line box in Fig.~\ref{fig:stream}), at a frequency of 100 kHz. From these recordings, the mixture fraction distributions $P(Z)$, as well as the flammability factor $F_f$ and the mean flammable mixture $\overline{Z}_{flam}$, are directly computed. At the same locations, the time-averaged mean and RMS data are used to construct MIST results. Figure~\ref{fig:flamma} shows the comparison between LES (left) and MIST (right). Note that the mesh of these maps is not regular as each location matches the position of a vertex of the unstructured LES grid.

\begin{figure}[ht!]
\centering
\includegraphics[width=0.48\textwidth]{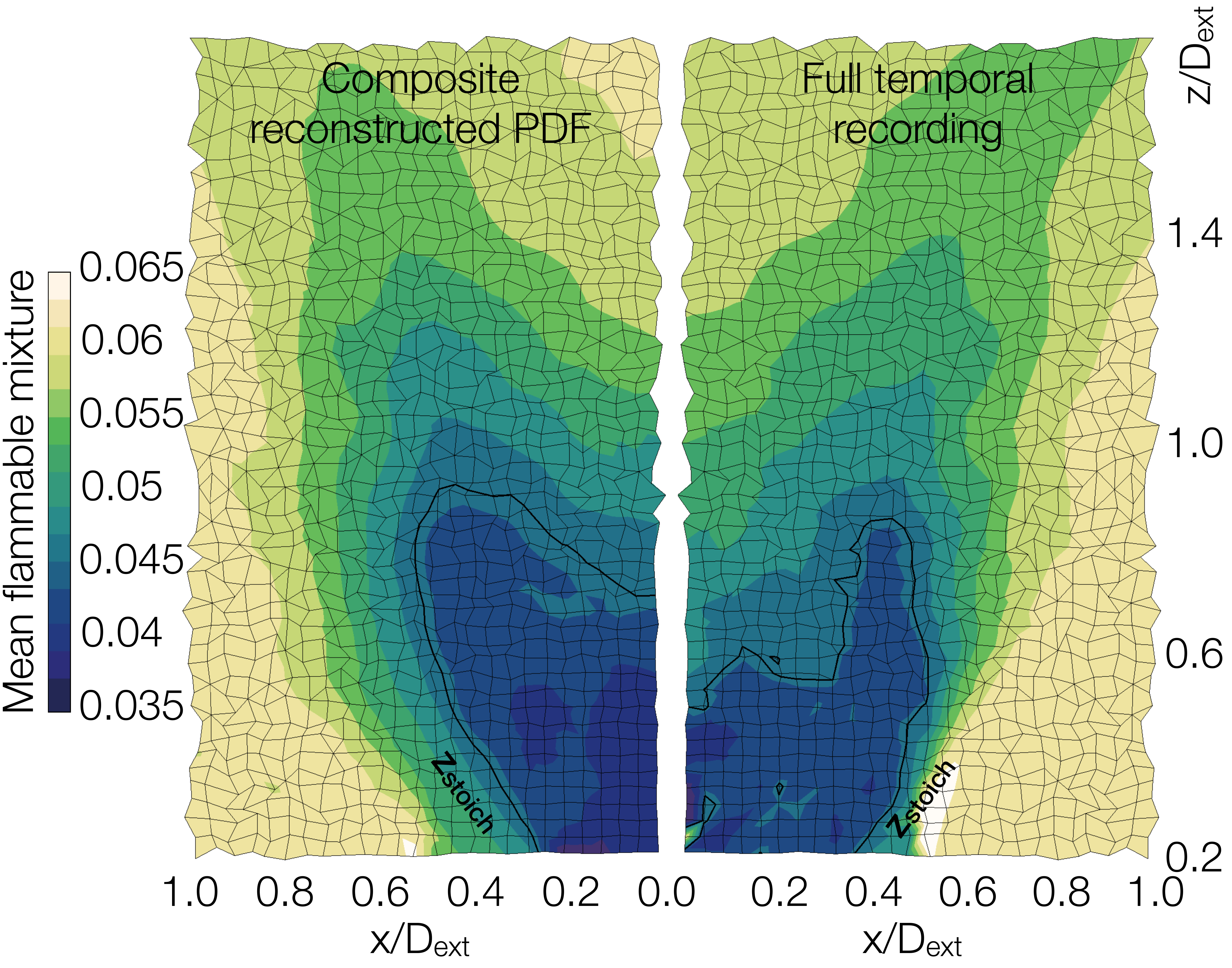}
\caption{Map of the mean flammable mixture fraction $\overline{Z}_{flam}$ obtained from MIST (left) and the LES (right). The black iso-contour indicates the position of stoichiometry.}
\label{fig:meanZ}
\end{figure}

The model is able to reproduce the main features of the flammability factor map but some discrepancies remain, up to an absolute error of 0.2 near the core of the methane jet. Predictions of the mean flammable mixture fraction $\overline{Z}_{flam}$ are also compared to the results directly extracted from the LES at the same locations. The comparison is provided in Fig.~\ref{fig:meanZ}, showing that the model provides a reasonably good estimation of $\overline{Z}_{flam}$. %The size of the central region with $\overline{Z}_{flam}$ above stoichiometry is overestimated by MIST but most of the iso-contour match the data extracted from temporal recordings.

\section{MIST algorithm overview}
\label{app:MISTflowchart}

\begin{figure}[ht!]
\centering
\includegraphics[width=0.5\textwidth]{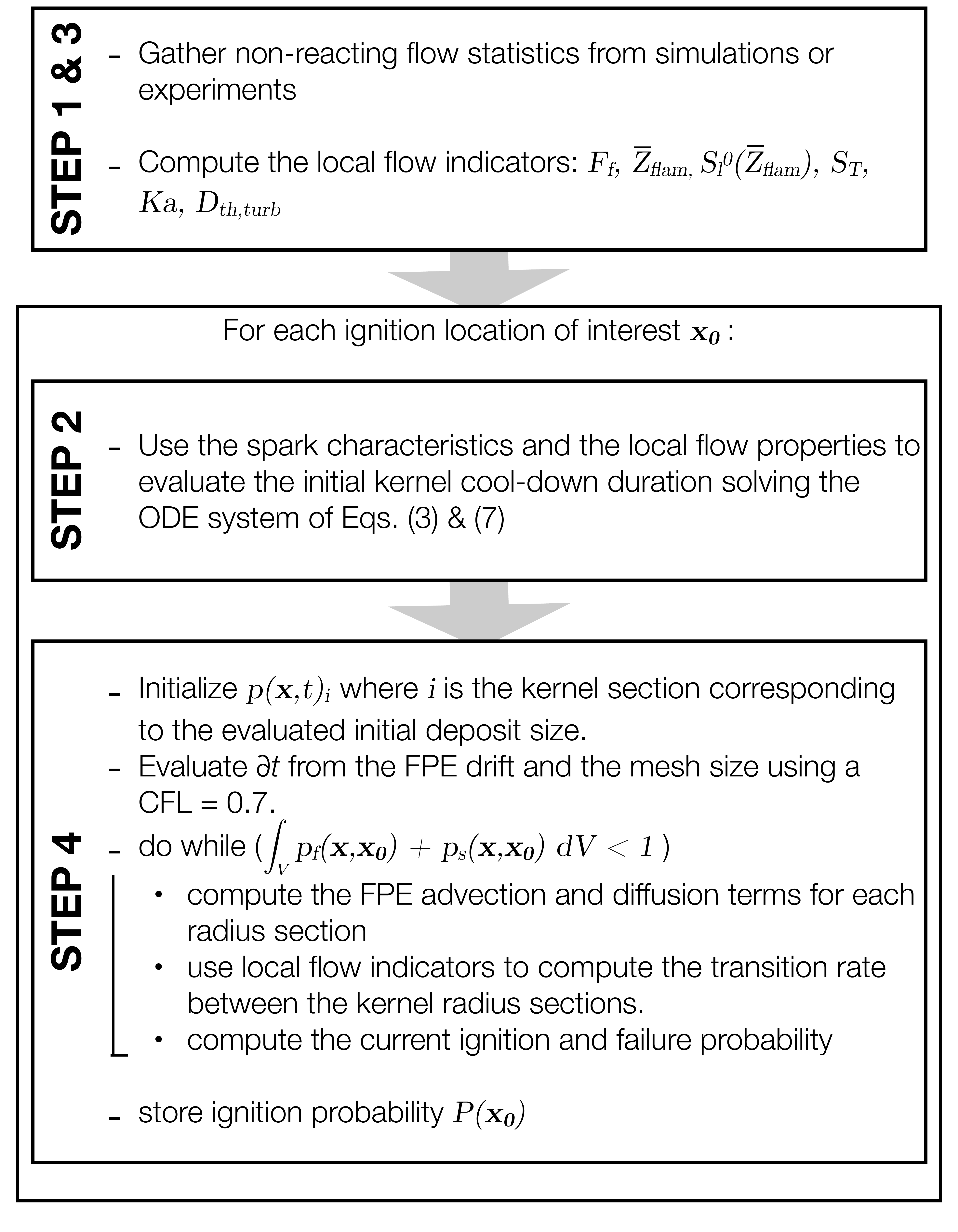}
\caption{Flowchart summarizing the successive steps and algorithm of MIST.}
\label{fig:flowchart}
\end{figure}

Figure~\ref{fig:flowchart} provides a flowchart description summarizing the main steps of MIST. Details of each steps are provided in the Section~\ref{sec_model}.

%Note that, although a similar direct comparison is not possible for the $SP$ case, it relies on the similar reconstruction of $P(z)$.

%\section{Validation of the velocity distribution}
%\label{app:velo}
%
%Although the assumption that the velocity follows a Gaussian distribution is not true from a turbulence point of view \cite{Pope:2000}, it is often made in deriving models because of the nice mathematical properties of the distribution. To evaluate the validity of this assumption in the present case, PDF of velocity components are constructed from temporal recording at the same locations used for the validation of the mixture distribution. The skewness factor, which can be viewed as a first order measure of the deviation from a Gaussian distribution, is showed in Fig.~\ref{} for all three velocity components.
%The results show that, departure from a Gaussian distribution can be locally significant. The radial velocity component exhibits strong positive skewness factor in the CRZ while the axial velocity skewness is significant in the IRZ. However, for the largest part of the map, the skewness factor stays in the [-1,1] range for which a Gaussian approximation is reasonable for the level accuracy desired for the approach.
%
%
%

%\listoftables
%\clearpage
%\input{./LIST/List_tab.tex}

%\listoffigures
%\clearpage
%\input{./LIST/List_fig.tex}

\end{document}

%% file: journaux-dec10.tex
% PLAIN FILE TO HAVE REFERENCES FOR COMBUSTION AND FLAME
% STYLE
\def\rtn{\par \noindent }
\def\pskip{\rtn }

%%%%% A %%%%%%%
\def\aa{ Acta Astronautica  }
\def\aam{ Adv. Appl. Mech.  }
\def\aas{ Atomization and Sprays  }

\def\aiaap{ AIAA Paper  }
\def\aj{ AIAA Journal  }
\def\aiaaj{ American Institute of Aeronautics and Astronautics Journal} 
\def\aiaap{ AIAA Paper}
\def\ajou{ Aeronaut. J.  }
\def\annr{ Ann. Rev. Fluid Mech. }
\def\taj{ Astrophys. J. } 	
\def\as{ Astron. and Astrophys. }
\def\autcf{Reprinted by permission of Elsevier Science from }
\def\autcff{\copyright ~the Combustion Institute }
\def\autjfm{Reprinted with permission by Cambridge University Press }

%%%%% B %%%%%%%
\def\bbpc{ Ber. Bunsenges. Phys. Chem. }

%%%%% C %%%%%%%
\def\canm{ Comm. Appl. Num. Meth. }
\def\ces{ Chem. Eng. Sci. }
\def\cf{ Combust. Flame }
\def\cfl{ Comput. Fluids }
\def\cip{ Comput. Phys. }
\def\cmame{ Comput. Methods Appl. Mech. Eng. }
\def\cpam{ Commun. Pure Appl. Math. }
\def\cpc{ Computer Phys. Communications }
\def\cras{ C. R. Acad. Sci. }
\def\cst{ Combust. Sci. Technol. }
\def\ctm{ Combust. Theor. Model. }
\def\ctrsp{ Proc. of the Summer Program }
\def\ctrarb{ Annual Research Briefs }
\def\ptrsla{Phil. Trans. R. Soc. London A }
\def\zmp{Z. Math. Phys. }
\def\jsiam{J. Soc. Indust. Appl. Math. }
\def\pla{Physics Letters A }
\def\ptrsa{Phil. Trans. R. Soc. A }

\def\csat{ Composites Science and Technology }              	% Composites Science and Technology
\def\csat{ Composites Sci. and Tech. }              			% Composites Science and Technology

%%%%% D %%%%%%%
\def\DHCRS{ Dynamics of Heterogeneous Combustion and Reacting Systems }

%%%%% E %%%%%%%
\def\ent{ Entropie }
\def\etme{ Eng. Turb. Modelling and Exp. } 
\def\exf{ Exp. Fluids }

%%%%% F %%%%%%%
\def\fd{ J. Fluid Dynamics }
\def\ftac{Flow Turbul. Combust.}

%%%%% I %%%%%%%
\def\icmf{ Int. Conf. Multiphase Flow }
\def\ijav{ Int. J. Acoust. Vib. }
\def\ijcfd{ Int. J. Comput. Fluid Dynamics }
\def\ijcm{ Int. J. Comput. Methods }
\def\ijhff{ Int. J. Heat Fluid Flow }
\def\ijmf{ Int. J. Multiphase Flow }
\def\ijmp{ Int. J. Modern Physics C }
\def\ijnme{ Int. J. Numer. Meth. Eng. }
\def\ijnmf{ Int. J.~Numer. Meth. Fluids }
\def\ijhmt{ Int. J.~Heat and Mass Transfer }
\def\ijts{ Int. J. of Therm. Sci. }
\def\iecf{Ind. Eng. Chem., Fundam.}
\def\ijck{ Int. J. Chem. Kinet. }                                      		% International Journal of Chemical Kinetics
\def\ijck{ International Journal of Chemical Kinetics }         	% International Journal of Chemical Kinetics
\def\ijhe{ Int. J. Hydrogen Energ. }

%%%%% J %%%%%%%

\def\ja{ J. Aircraft }
\def\jacic{ J. Aeropace Comput. Inform. Comm. }
\def\jaes{ J. Aeronaut. Sci. }
\def\jam{ SIAM J. Appl. Math. }
\def\jame{ J. Appl. Mech. }
\def\jamp{ J. Appl. Phys. }
\def\jars{ J.~American Rocket Society }
\def\jas{ J. Atmos. Sci. }
\def\jasa{ J. Acous. Soc. Am. }
\def\jchp{ J. Chem. Phys. }
\def\jcht{ J. Chem. Thermodynamics }
\def\jcp{ J.~Comput. Phys. }
\def\je{ J. Energy }
\def\jep{ J. Eng. Gas Turb. and Power }
\def\jfe{ J. Fluids Eng. }
\def\jfm{ J.~Fluid Mech. }
\def\jfs{ J. Fluids Struct. }
\def\jht{ J. Heat Trans. }
\def\jie{ J. Inst. Energy }
\def\jmta{ J. M\'ec. Th\'eor. Appl. }
\def\jpp{ J.~Prop.~Power }
\def\jrnbs{ J. Res. Natl. Bur. Stand. }
\def\jsc{ J. Sci. Comput. }
\def\jsr{ J. Spacecrafts and Rockets }
\def\jsv{ J.~Sound Vib. }
\def\jssc{ SIAM J. Sci. Stat. Comput. }
\def\jt{ J.~Turb. }
\def\jtht{ J. Thermophysics and Heat Trans. }
\def\jtu{ J. Turbomach. }
\def\jpca{J. Phys. Chem. A}
\def\jpdap{ Journal of Physics D: Applied Physics }          	% Journal of Physics D: Applied Physics
\def\jpdap{ J. Phys. D: Appl. Phys. }          				% Journal of Physics D: Applied Physics
\def\jegtp{J. Eng. Gas Turb. Power}

%%%%% M %%%%%%%
\def\moc{ Math. Comp. }
\def\mwr{ Mon. Weather Rev. }

%%%%% N %%%%%%%
\def\ned{ Nuclear Eng. and Design }

%%%%% P %%%%%%%
\def\paa{ Prog. in Astronautics and Aeronautics }
\def\pas{ Prog. Aerospace Sci. }
\def\pcfd{ Prog. Comput. Fluid Dynamics }
\def\pci{ Proc. Combust. Inst. }
\def\pcp{ Prog. Comput. Phys. }
\def\pecs{ Prog. Energy Comb. Sci. }
\def\pf{ Phys. Fluids }
\def\pime{ Proc. Instn. Mech. Engrs. }
\def\plms{ Proc. London Math. Soc }
\def\ppsc{ Particle and Particle Systems Characterization }
\def\prl{ Phys. Rev. Lett. }
\def\prsl{ Proc. R. Soc. Lond. }
\def\prsla{ Proc. R. Soc. Lond. A }
\def\pt{ Powder Technology }
\def\pep{ Propellants, Explosives, Pyrotechnics }          	% Propellants, Explosives, Pyrotechnics

%%%%% Q %%%%%%%
\def\qjmam{ Q. J. Mech. Appl. Math. }

%%%%% R %%%%%%%
\def\ra{ La Rech. A\'{e }rospatiale }
\def\rpa{ Rev. Phys. Appl. }

%%%%% T %%%%%%%
\def\tcfd{ Theoret. Comput. Fluid Dynamics }
\def\tcsme{ ASME Trans. }
\def\ti{ Technique de l'Ing\'enieur }
\def\tsfp{ Turb. Shear Flow Phenomena }

%%%%% W %%%%%%%
\def\WSSCI{ WSS/CI }

\def\Litem{\par\noindent }
\font\smc=cmcsc10